\documentclass[letterpaper, journal]{IEEEtran}
\usepackage{amsmath,amsfonts}
\usepackage{algorithmic}
\usepackage{algorithm}
\usepackage{array}
\usepackage[caption=false,font=normalsize,labelfont=sf,textfont=sf]{subfig}
\usepackage{textcomp}
\usepackage{stfloats}
\usepackage{url}
\usepackage{verbatim}
\usepackage{graphicx}
\usepackage{cite}
\usepackage{multirow}
\usepackage[table,xcdraw]{xcolor}
\usepackage{booktabs}
\usepackage{adjustbox}
\usepackage{orcidlink}

\begin{document}

\title{FedSlate: A Federated Deep Reinforcement Learning Recommender System\\
{\footnotesize \textsuperscript{*}This article has been accepted for publication in IEEE Transactions on Emerging Topics in Computational Intelligence (TETCI). \copyright~IEEE. Personal use is permitted, but republication/redistribution requires IEEE permission.}}


\author{\thanks{\IEEEauthorrefmark{1} This is to indicate the equal contribution}\thanks{\IEEEauthorrefmark{2} This is to indicate the corresponding author}  Yongxin Deng\orcidlink{https://orcid.org/0009-0003-4352-1308}\IEEEauthorrefmark{1}, Xihe Qiu\orcidlink{https://orcid.org/0000-0003-4024-925X}\IEEEauthorrefmark{1}\IEEEauthorrefmark{2}, Xiaoyu Tan\orcidlink{https://orcid.org/0000-0003-3555-7143}\IEEEauthorrefmark{1} and Yaochu Jin\orcidlink{https://orcid.org/0000-0003-1100-0631}~\IEEEmembership{Fellow,~IEEE}
\thanks{Yongxin Deng,  Xihe Qiu (email:qiuxihe@sues.edu.cn) are with the School of Electronic and Electrical Engineering, Shanghai University of Engineering Science, Shanghai, China} \thanks{Xiaoyu Tan is with the INFLY TECH (Shanghai) Co., Ltd. Shanghai, China }\thanks{Yaochu Jin is with the School of Engineering, Westlake University, Hangzhou, China}}

\markboth{Journal of \LaTeX\ Class Files,~Vol.~14, No.~8, August~2021}%
{Shell \MakeLowercase{\textit{et al.}}: A Sample Article Using IEEEtran.cls for IEEE Journals}


\maketitle

\begin{abstract}
Reinforcement learning methods have been used to optimize long-term user engagement in recommendation systems. However, existing reinforcement learning-based recommendation systems do not fully exploit the relevance of individual user behavior across different platforms. One potential solution is to aggregate data from various platforms in a centralized location and use the aggregated data for training. However, this approach raises economic and legal concerns, including increased communication costs and potential threats to user privacy. To address these challenges, we propose \textbf{FedSlate}, a federated reinforcement learning recommendation algorithm that effectively utilizes information that is prohibited from being shared at a legal level. We employ the SlateQ algorithm to assist FedSlate in learning users' long-term behavior and evaluating the value of recommended content. We extend the existing application scope of recommendation systems from single-user single-platform to single-user multi-platform and address cross-platform learning challenges by introducing federated learning. We use RecSim to construct a simulation environment for evaluating FedSlate and compare its performance with state-of-the-art benchmark recommendation models. Experimental results demonstrate the superior effects of FedSlate over baseline methods in various environmental settings, and FedSlate facilitates the learning of recommendation strategies in scenarios where baseline methods are completely inapplicable. Code is available at \textit{https://github.com/TianYaDY/FedSlate}.
\end{abstract}


\section{Introduction}
\IEEEPARstart{G}{iven} the significant influence of users' instant reactions to recommended content on their future behavior, research in the realm of content recommendation systems has increasingly adopted deep reinforcement learning (DRL) strategies. These strategies aim to optimize the balance between immediate user engagement and long-term retention \cite{RN1,RN16,RN15}. Additionally, recent advancements in advertising recommendation systems have integrated reinforcement learning to achieve a balance between generating ad revenue and minimizing negative user experiences \cite{RN24}. However, these systems often overlook an essential aspect: user behavior is not isolated but affected by recommendations from various platforms, indicating interdependencies between a user's activities across different services. An obvious solution to leverage these behavioral correlations is to consolidate user data from multiple sources for the development of a unified model. Nevertheless, the introduction of stringent data privacy regulations, such as the General Data Protection Regulation (GDPR) in the European Union, the Personal Data Protection Act (PDPA) in Singapore, and the California Consumer Privacy Act (CCPA) in the United States, raises significant legal challenges. Directly employing user data could infringe upon privacy rights \cite{RN18,RN22}. Additionally, the communication overhead \cite{RN21,RN20,RN19} poses a barrier to the straightforward implementation of centralized learning approaches. To circumvent these obstacles while harnessing the coherence in user behavior across platforms, we advocate for the adoption of federated learning (FL) techniques \cite{DBLP:conf/aistats/McMahanMRHA17} to refine reinforcement learning-based recommendation algorithms.\par

In numerous contexts, the correlation between user behaviors yields substantial benefits. Within the financial sector, for instance, customers often engage with both stock trading and online payment services. A collaborative model trained by these services can effectively ascertain a customer's risk profile, investment patterns, and spending behaviors, thereby facilitating tailored financial product recommendations that meet individual needs. Similarly, in the realm of online advertising, user interactions with multiple platforms reveal interconnected behaviors. A user may explore health foods on a social network while simultaneously shopping for wellness products on an e-commerce site. Utilizing FL, platforms can collectively develop a model that captures the user's interests and buying inclinations, enhancing the precision of targeted advertising. Consequently, a FL-based recommender system offers a potential approach that optimizes user privacy and enhances long-term value (LTV). However, previous research on FL in recommendation systems has primarily focused on conventional algorithms with simple averaging of local model outputs \cite{yang2020federated,jalalirad2019simple,muhammad2020fedfast,tan2020federated}. These approaches often fail to account for user heterogeneity \cite{imran2023refrs}, which is particularly problematic given the inherently diverse and imbalanced nature of user preferences, potentially resulting in systems that inadequately capture long-term user interests.\par

Recommendation systems commonly adopt a slate-based approach, wherein multiple items are simultaneously suggested to the user, allowing them to select and view their preferred item. This presents a significant challenge for the direct application of reinforcement learning (RL) due to the large action space involved. SlateQ \cite{RN1} is a recommendation algorithm that utilizes the slate decomposition technique to address the challenge of recommending multiple items, known as a recommendation slate, to users. It effectively resolves the issue of the large action space faced by previous RL recommendation algorithms. However, SlateQ can only be deployed separately on different platforms and cannot be easily extended to the scenario of joint deployment across multiple platforms. As a result, it fails to leverage the correlations between user behaviors on different platforms. To address this problem, we propose FedSlate, assuming that a user's behavior and response on one platform can be influenced by the recommended content from other platforms. Furthermore, certain influences are regarded as ``inaccessible'' to specific platforms. Specifically, we assume that certain agents are unable to directly receive rewards (even though the rewards exist), and these agents cannot make decisions based solely on their own information.\par

Our FedSlate algorithm incorporates both local and global models. However, unlike the adaptive personalized federated learning (APFL) algorithm \cite{RN4}, we do not blend the local and global models proportionally. Instead, we adopt a method similar to ``Q-value sharing'' \cite{RN7}, where the local models generate Q-values that are passed as inputs to the central server. The central server then computes the global Q-values used for content recommendation selection. FedSlate can be divided into the following stages. First, dedicated agents on each platform calculate local Q-values based on observed user states and candidate recommendation content states. They transmit these values to the central server. Next, the central server collects the received local Q-values as inputs to the global Q-network and calculates the corresponding global Q-values for each local agent. Finally, the central server distributes the Q-values to the respective agents, and the local agents make policy selections based on the received Q-values. It is important to note that each agent is unaware of the Q-network parameters of other agents. The central server calculates the global Q-values as many times as there are local agents since each agent has its own distinct ``local Q-values''. These stages do not include the process of updating the local and global Q-networks. In FedSlate, we iteratively update the global network and local network based on the outputs of the global Q-network and the rewards obtained by an agent in its environment. During this process, the global network is updated multiple times, while the local network is updated only once.\par

We summarize the main contributions into threefold:
\begin{enumerate}

\item We propose FedSlate, a novel algorithm that integrates SlateQ methodology with federated learning principles, enabling effective monitoring of long-term user behavior patterns while incorporating cross-platform recommendation effects.

\item We design an innovative update mechanism that implements sequential updating protocol, involving multiple updates of the global network while maintaining single updates for local networks, thereby enhancing both algorithmic efficiency and output quality.

\item We demonstrate FedSlate's effectiveness in developing recommendation strategies using data not directly accessible to local agents, addressing practical challenges in privacy-restricted scenarios.

\end{enumerate}

In Section \ref{Related Work}, we review relevant prior research related to our FedSlate algorithm. Subsequently, in Section \ref{Problem Definition}, we present the problems that FedSlate aims to resolve. Section \ref{Our FedSlate Method} describes the intricacies of the FedSlate algorithm. Finally, in Section \ref{Experimental Setup}, we deploy RecSim \cite{RN26} to establish a simulation environment for evaluating recommender systems and to examine the efficacy of the FedSlate algorithm in this context.
\section{Related Work}
\label{Related Work}
\subsection{Recommender Systems}
Recommender systems are a critical type of information filtering system that leverages user preference and behavior analysis to provide personalized suggestions \cite{lu2015recommender}. These systems are extensively applied in various sectors, including e-commerce, social media, and entertainment platforms like music and video streaming, aiming to enhance content discovery and improve the overall user experience. Supervised learning (SL) techniques are commonly utilized in these systems to perform predictive and recommendatory functions, relying on patterns and rules derived from labeled training data. In this realm, training datasets, which include users' historical interactions and their corresponding feedback or ratings, are instrumental. Among the prevalent SL-based methodologies for recommender systems, collaborative filtering (CF) stands out \cite{su2009survey,koren2021advances}. User-based CF \cite{thorat2015survey} suggests items by identifying similarities between users' past behaviors, positing that users with comparable preferences are likely to be interested in similar items. Conversely, item-based CF \cite{wang2006unifying}, recommends based on item similarities. These CF methods are favored for their simplicity and proven effectiveness. Content-based recommendations \cite{pazzani2007content} represent another widespread SL approach, employing item characteristics and user preferences to formulate suggestions. For example, a movie recommendation system may use a content-based method to recommend movies by considering aspects such as genre, actors, and directors, thus predicting a user's potential interests. Furthermore, SL-based recommender systems may integrate various machine learning models, including decision trees \cite{gershman2010decision}, support vector machines (SVM) \cite{oku2006context}, and deep neural networks (DNN) \cite{gupta2020architectural}. These models enhance the recommendation process by adapting to diverse dataset features and characteristics while learning from users' preferences and behavior patterns during training.

SL-based recommender systems have shown proficiency in short-term prediction tasks; however, incorporating reinforcement learning (RL) has been identified as critical for long-term prediction challenges \cite{RN27, RN28, tan2023adaptive}. RL, a machine learning paradigm, seeks to establish optimal behavioral policies through environmental interactions \cite{9904958,kaloev2021experiments} and distinguishes itself by concentrating on goal-directed decision-making, employing a trial-and-error process with a rewards system. This method excels in scenarios requiring foresight, such as financial investment \cite{deng2016deep} and complex planning, due to its adeptness at managing delayed rewards. In the medical domain, where uncertainty and dynamic conditions are prevalent, like ventilator management \cite{QIU2022107689, CHEN202247}, RL's attributes prove exceptionally beneficial. Within RL-based recommendations, there are ``model-based'' and ``model-free'' methods; our focus is on the ``model-free'' category, which is straightforward to implement and yields superior long-term performance. Before this approach, DRN \cite{RN29} implemented a deep Q-network (DQN) to create user profiles and an activity score. Subsequently, the social attentive deep Q-network (SADQN) \cite{RN30} enhanced DQN with an attention mechanism to leverage social influences. Diverging from these, our contribution, FedSlate, leverages cross-platform user performance similarities. While some studies have adopted policy gradient methods, such as the Monte Carlo-based REINFORCE algorithm for large-scale recommendation environments \cite{RN32}, the SlateQ algorithm \cite{RN1} is particularly influential in our approach. It decomposes the Q-value of a recommendation slate into individual item Q-values, effectively managing extensive action spaces and offering three item-wise Q-value-based selection strategies. Notwithstanding, existing RL-based systems primarily optimize for single-platform performance, neglecting the multiplatform influences on real-world users. To address this gap, we propose the incorporation of FL paradigms into RL-based recommendation systems.

\subsection{Federated Learning}
FL involves training statistical models directly on devices to develop a joint model capable of generating data across distributed nodes \cite{DBLP:conf/aistats/McMahanMRHA17}. Traditional distributed learning methods generally presuppose that local data samples are independently and identically distributed (IID); however, FL typically operates under the premise that data among clients is non-IID \cite{RN8,sattler2019robust}. Prior research has primarily utilized RL to enhance FL's performance \cite{RN35}. For instance, \cite{RN35} employs a Deep Q-Learning (DQL) strategy to select devices for participation in successive communication rounds, thus minimizing the number of required rounds. Related work can be found in \cite{RN36, RN38}. In contrast, \cite{RN37} applies a deep reinforcement learning (DRL) technique to modulate the CPU frequency of faster devices within an FL training cohort, balancing energy efficiency with training velocity. Distinct from these studies, our FedSlate algorithm ``federates'' DRL rather than simply applying RL to optimize FL. Concerning federated recommender systems, current studies \cite{yang2020federated,jalalirad2019simple,muhammad2020fedfast,tan2020federated}, have mainly adapted FL for traditional recommendation frameworks, setting them apart from our DRL-centric FedSlate approach. The most significant influence on our work is Federated deep Reinforcement Learning (FedRL) \cite{RN7}, which introduces an innovative DRL paradigm to collaboratively construct high-quality models for agents. Our approach adopts the ``Q-value sharing'' concept from FedRL to monitor long-term user behaviors and evaluate the impact of recommended content from various platforms on a user's actions within a specific platform, thus tackling the complexities of applying RL recommendation algorithms in a multi-platform context. While FedRL advocates for a decentralized scheme, our FedSlate algorithm is designed with centralization in mind but can also be adapted to a decentralized format for practical applications. Please note that although our framework involves a central party, this central party does not have full transparency in accessing information from various platforms. The information that the central party can obtain is strictly limited to the Q-values generated by each platform, rather than sensitive user privacy data. Similarly, the data transmitted by the central party to each party is confined to the processed Q-values. In the FL process, such information exchange is inevitable, and there is essentially no difference between transmitting it to discrete parties and transmitting it to a central party, especially when communication is with a trusted central entity that can ensure data protection. We introduced a centralized architecture to better construct an algorithmic abstraction, thereby formally aligning with FL's pioneering FedAvg algorithm \cite{DBLP:conf/aistats/McMahanMRHA17}. In reality, the central party is an abstract rather than an instantiated concept, meaning that any platform instance can serve as the central party.

\section{Problem Definition}
\label{Problem Definition}
In this section, we introduce an augmented Markov Decision Process (MDP) model tailored for the single-user, multi-platform context. This model captures the dynamics wherein platforms employ recommendation systems to curate slates of content. Users engage with these slates by selecting an item—or opting out—and post-consumption, decide whether to seek additional recommendations or end their session. It is important to note that users may transition across platforms in pursuing content that piques their interest, a pattern that closely reflects real-world user behavior. User responses to content are multifaceted, encompassing metrics such as browsing duration, ``likes'', and comments. However, for the sake of a generalized model, we limit our focus to user engagement as the singular metric of reward. Subsequently, we outline the assumptions underpinning our problem, some aligning with the SlateQ framework \cite{RN1} and others specific to the single-user, multi-platform scenario. We conclude this section by detailing a precise formalization of our federated reinforcement learning recommendation problem.\par

\subsection{An Extended MDP Model for Slate Recommendation}
\label{sec:extmdp}
The recommendation and user interaction behaviors within a single platform are aptly modeled by an MDP, characterized by states $S$, actions $A$, a reward function $R$, a transition kernel $P$, and a discount factor $\gamma$ \cite{RN1}. We will now elucidate the critical elements of this model:
\begin{itemize}
\item The state $\mathcal{S}$ implements the user's condition, comprising both observable attributes such as age, gender, and self-reported interests, and historical interactions including prior browsing activity and responses to earlier recommendations.
\item Action $\mathcal{A}$ encompasses all potential recommendation arrays; upon generating a collection of $\mathcal{I}$ items, the system is tasked with curating a subset of $k$ items to form the user's slate, denoted by ${A\subseteq \mathcal{I}\quad s.t.\left| A \right|=k}$, where $k$ is the pre-defined slate size.
\item The transition probability $P(s{}'| s,A)$ quantifies the likelihood of migrating from state $s$ to state $s{}'$ subsequent to action $A$.
\item The reward $R(s,A)$ assesses the anticipated user engagement with the slate $A$, serving as an index of the user's interaction with the recommended items.
\end{itemize} 
The value function or Q-function of the policy $\pi: S\to A$ that the agent takes after observing state $s$ is given by the following equation:\par
\begin{equation}
V^{\pi}\left ( s \right )  = R\left (s,\pi \left (s\right )\right )+\gamma \sum _{s{}'\in S }P\left (s{}'| s,\pi \left (s\right )\right )V^{\pi }\left(s{}'\right )
\label{eq.1}
\end{equation}
\begin{equation}
Q^{\pi}\left ( s,A \right )  =R\left (s,A \right )+\gamma \sum _{s{}'\in S }P\left (s{}'| s,A\right )V^{\pi }\left(s{}'\right )
\label{eq.2}
\end{equation}\par
The optimal policy $ \pi ^{\ast } $ maximizes the expected value $ V\left(s\right)$. Therefore, we focus on the following expression:\par
\begin{equation}
V^{\ast}\left ( s \right )  =\mathop{max}_{A\in \mathcal{A} }^{} R\left (s,A\right )+\gamma \sum _{s{}'\in S }P\left (s{}'| s,A\right )V^{\ast }\left(s{}'\right )
\label{eq.3}
\end{equation}
\begin{equation}
Q^{\ast}\left ( s,A \right )  =R\left (s,A\right )+\gamma \sum _{s{}'\in S }P\left (s{}'| s,A\right )V^{\ast }\left(s{}'\right )
\label{eq.4}
\end{equation}
\par

Within this framework, the optimal policy $\pi^{\ast}$ is defined such that $\pi^{\ast}(s) = \underset{A \in \mathcal{A}}{\mathrm{argmax}} , Q^{\ast}(s,A)$. \par

To better reflect real-world usage patterns, we consider scenarios where users interact with multiple platforms simultaneously. Our algorithm is demonstrated in this multi-platform context.

However, the standard MDP framework proves insufficient for modeling cross-platform user behavior. To address this limitation, we propose an enhanced MDP framework encompassing platforms A and B, where individual MDPs operate independently on each platform. Concretely, this involves discrete recommendation agents for each platform (agent $\alpha$ for platform A and agent $\beta$ for platform B), with corresponding user states ($S_{\alpha}$ and $S_{\beta}$), agent actions ($A_{\alpha}$ and $A_{\beta}$), transition probabilities ($P_{\alpha}(s' | s, A)$ and $P_{\beta}(s' | s, A)$), rewards ($R_{\alpha}$ and $R_{\beta}$), and Q-functions ($Q_{\alpha}$ and $Q_{\beta}$).

\subsection{Necessary Assumptions}
In our federated reinforcement learning recommendation problem, we make the following assumptions:
\begin{itemize}
    \item \textbf{A1:} A user selects only one item at a time (or may choose not to select, represented as $\bot$ for a null item).
    \item \textbf{A2:} Transitions depend solely on the selection. Specifically, the user's state changes, and a reward (user engagement) is generated only when the user consumes item $i$. Additionally, while a user engages with a platform, the states of other platforms remain frozen.
    \item \textbf{A3:} There is interconnectedness between the user's behaviors on different platforms, and the impact of a single platform on the user is ``cross-platform''.
    \item \textbf{A4:} Only the output values of $Q_{\alpha}$ and $Q_{\beta}$ are shared for learning the joint policy $\pi^{\ast}_{fed}$. Other information, including transitions $D_{\alpha}=\left \{ \left \langle  s_{\alpha},A_{\alpha},s{}'_{\alpha},r_{\alpha} \right \rangle  \right \} $ and $D_{\beta}=\left\{\left\langle s_{\beta},A_{\beta}\right \rangle\right\}$, is locally visible only.
\end{itemize} 
\textbf{A1} is the original assumption in SlateQ (\textbf{Single choice (SC)}). \textbf{A2} extends the \textbf{Reward/transition dependence on selection (RTDS)} assumption in SlateQ to the multi-platform scenario. \textbf{A3} and \textbf{A4} are specific assumptions for the single-user multi-platform context. \textbf{A3} is the most important premise of our method, which is intuitive and easily acceptable.
\textbf{A4} ensures that information about the user on different platforms is not leaked (we believe that even for the same user, information should not be shared between platforms without user authorization).
\subsection{Recommendation Problem}
After extending the original MDP model, we can formally define our recommendation problem based on \textbf{A1}-\textbf{A4}. The existing platforms A and B take turns randomly recommending slates to customers and recording transitions. This results in a series of transitions $D_{\alpha}$ for agent $\alpha$ and transitions $D_{\beta}$ for agent $\beta$, where $D_{\alpha}$ and $D_{\beta}$ are one-to-one correspondence. Our goal is to learn a joint policy $\pi^{\ast}$ that, based on $s_{\alpha}$ and $s_{\beta}$, maximizes the lifetime value (LTV) across all platforms. It should be noted that platform B does not record the user's response, i.e., $D_{\beta}$ does not contain $r_{\beta}$, which deviates from the previous MDP model. We adopt this setting because certain platforms may not have direct access to user feedback on recommended content (although the platform's impact on the user is real). User preferences or aversions may be reflected on other platforms. We aim to demonstrate the friendliness of FedSlate towards these ``unavailable feedback'' platforms—even if a platform cannot directly update its recommendation strategy based on user feedback, it can still benefit from performance improvements in the federation.

\section{Our FedSlate Method}
\label{Our FedSlate Method}
In this section, we will provide a detailed description of our FedSlate method. We utilize the SlateQ algorithm to assist us in evaluating the value of recommended content and tracking user feedback over the long term. Therefore, we will begin by briefly introducing the original SlateQ algorithm. Subsequently, we will present the components of our algorithm, followed by a description of the specific details of the algorithm. Lastly, we will propose an extended version of our algorithm to address situations where team rewards are excessively sparse.

\subsection{SlateQ Algorithm}
In Section \ref{sec:extmdp}, we presented the MDP model for the recommendation problem. For a single-user, single-platform MDP, the original SlateQ algorithm aims to find an optimal policy $\pi^{\ast}$ that satisfies $ \pi^{\ast}\left(s\right )=argmax_{A\in \mathcal{A} }Q^{\ast}\left (s,A\right )$. However, in slate recommendation problems, the action space grows exponentially, imposing substantial computational and temporal complexity. Specifically, selecting and ordering $k$ items from a set of $\mathcal{I}$ items into an ordered slate results in an action space of cardinality $P(\mathcal{I},k) = \frac{|\mathcal{I}|!}{(|\mathcal{I}|-k)!}$, where $P(\mathcal{I},k)$ denotes the number of $k$-permutations of $\mathcal{I}$. \par

To address this issue, SlateQ decomposes $Q^{\pi}\left ( s,A\right )$ and represents slate-level Q-values as item-level Q-values $\overline{Q}^{\pi}\left ( s,i\right )$, significantly reducing the agent's action space. The decomposition is achieved using the following formula:
\begin{equation}
    Q^{\pi}\left ( s,A\right ) = \sum_{i\in A}P\left(i|s,A\right)\overline{Q}^{\pi}\left ( s,i\right )
\end{equation}
The decomposed Q-values can be updated using a simple Temporal Difference (TD) method:
\begin{equation}
\overline{Q}^{\pi} (s,i)\gets\alpha(r+\gamma\sum_{j\in A'}P(j|s',A')\overline{Q}^{\pi} ( s',j ))+(1-\alpha)\overline{Q}^{\pi}\left ( s,i\right )
\label{eq6}
\end{equation}
To fully satisfy the requirements of Q-learning, it is only necessary to introduce the usual maximization step:
\begin{equation}
\overline{Q} ( s,i )\gets\alpha(r+\gamma \mathop{max}_{A'\in \mathcal{A} }  \sum_{j\in A'}P(j|s',A')\overline{Q} ( s',j ))+(1-\alpha)\overline{Q}\left ( s,i\right )
\label{eq7}
\end{equation}
Please note that SlateQ assumes the user choice model $P(i|s,A)$ is known. Models such as multinomial logit model (MNL) \cite{louviere2000stated}, conditional logit model (CL), and cascade model \cite{DBLP:conf/kdd/Joachims02, DBLP:conf/wsdm/CraswellZTR08} can easily be learned using user response data, and this does not depend on LTV.\par
SlateQ offers multiple strategies to select recommended slates based on item Q-values. We consider the trade-off between computational and time resources and the effectiveness of the strategy, and we will only provide a detailed explanation of the Greedy optimization approach. A simple approach to utilizing item Q-values in slate construction is to use the Q-values as item scores, sort the items in descending order of scores, and select the top $k$ items to form the slate. However, this approach, known as the Top-$k$ method, fails to capture the influence of the first $L-1$ items on the $L$th slot (for $1<L\le k$). Greedy optimization differs from the aforementioned method as it updates the item scores based on the current partial slate. For example, given $A'=\{i_{(1)},...,i_{(L-1)}\}$ of size $L-1<k$, the $L$th item is selected based on the maximum marginal value it provides:
\begin{equation}
\mathop{argmax}_{i\notin A'}\frac{v(s,i)\overline{Q}(s,i)+\textstyle\sum_{l<L}v(s,i_{(l)})\overline{Q}(s,i_{(l)})}{v(s,i)+v(s,\perp)+\textstyle\sum_{l<L}v(s,i_{(l)})}
\label{eq8}
\end{equation}\par

\subsection{The FedSlate Algorithm}
\label{method:fedslate}
In this section, we will introduce our FedSlate algorithm in a bottom-up manner, starting from some necessary components. \par
\paragraph{\textbf{Basic Q Networks}}
We establish two Q networks, denoted as $Q_{\alpha}\left (s_{\alpha};{\theta}_{\alpha}\right )$ and $Q_{\beta}\left (s_{\beta};{\theta}_{\beta}\right )$, for agents $\alpha$ and $\beta$ respectively. Here, ${\theta}_{\alpha}$ and ${\theta}_{\beta}$ represent the parameters of the Q networks. It should be noted that both Q networks have the same structure as the Q network in the original SlateQ algorithm. Their output is a tensor of the same size as the number of candidate documents $\mathcal{I}$ (equivalent to the set $\overline{Q}(s,i), i \in \mathcal{I}$). However, we do not directly use their output values for slate recommendations. Instead, we use the output values as inputs for the federated Q network.\par

\paragraph{\textbf{Federated Agent}}
To exploit information from both Platform A and Platform B, we introduce a third agent, referred to as agent $fed$, which receives the output values of $Q_{\alpha}$ and $Q_{\beta}$.\par

Within agent $fed$, we construct a simple multi-layer perceptron (MLP) module, also known as the federated Q network mentioned earlier, denoted as $Q^{f}$. This network utilizes the output of the two basic Q networks to derive Q values specifically used for slate selection. When $Q^{f}$ is employed for slate content selection, it differs depending on whether it is used by agent $\alpha$ or agent $\beta$. Specifically, agent $\alpha$ and agent $\beta$ have their own respective output values from $Q^{f}$, denoted as $Q_{\alpha}^{f}$ and $Q_{\beta}^{f}$, defined as follows:\par
\begin{equation}
    Q_{\alpha}^f(\cdot;\theta_{\alpha},\theta_{\beta},\theta_{f})=MLP([Q_{\alpha}(s_{\alpha};\theta_{\alpha})|Q_{\beta}(s_{\beta};\theta_{\beta});\theta_{f})
\end{equation}
\begin{equation}
    Q_{\beta}^f(\cdot;\theta_{\alpha},\theta_{\beta},\theta_{f})=MLP([Q_{\beta}(s_{\beta};\theta_{\beta})|Q_{\alpha}(s_{\alpha};\theta_{\alpha});\theta_{f})
\end{equation}\par
Where $\theta_{f}$ represents the parameters of the MLP, and $[\cdot|\cdot]$ denotes the concatenation operation. In the above equation, we utilize the first position of the MLP input to represent ``one's own Q value'', while the second position represents ``Q values that do not belong to oneself''.\par

During the actual process of slate recommendation and Q network update, we fix the parameters $\theta$ of the Q network for agents on the platform where the user is not present. We treat the output of the Q network as a constant to ensure the stability of our algorithm during the learning phase,\par
\begin{equation}
    Q_{\alpha}^f(\cdot,C_{\beta};\theta_{\alpha}, \theta_{f})=MLP([Q_{\alpha}(s_{\alpha};\theta_{\alpha})|C_{\beta});\theta_{f})
    \label{eq11}
\end{equation}
\begin{equation}
    Q_{\beta}^f(\cdot ,C_{\alpha};\theta_{\beta}, \theta_{f})=MLP([Q_{\beta}(s_{\beta};\theta_{\beta})|C_{\alpha});\theta_{f})
        \label{eq12}
\end{equation}\par
Where $C_{\alpha}=Q_{\alpha}(s_{\alpha};\theta_{\alpha})$ and $C_{\beta}=Q_{\beta}(s_{\beta};\theta_{\beta})$ are the fixed outputs of the basic Q networks that we mentioned earlier\footnote{Please note that, for the sake of brevity in expression, we will not differentiate between the Q networks themselves and their output values in the following text. We will uniformly use $Q$ to represent them.}.\par
Agent $fed$ is responsible for training the Q-networks. During the training phase, agent $fed$ sequentially receives $Q_{\alpha}$ and $Q_{\beta}$ and updates the corresponding networks. In order to minimize the error, we use the Huber loss to define the loss functions $L_{\alpha}(\theta_{\alpha},\theta_{f})$ and $L_{\beta}(\theta_{\beta},\theta_{f})$ for agents $\alpha$ and $\beta$ respectively:
\begin{equation}
\mathcal{L}_{\delta}(a)=\left\{\begin{array}{ll}
\frac{1}{2} a^{2} & \text { for }|a| \leq \delta , \\
\delta \cdot\left(|a|-\frac{1}{2} \delta\right), & \text { otherwise. }
\end{array}\right.
\end{equation}
\begin{equation}
L_{\alpha}(\theta_{\alpha},\theta_{f})=\frac{1}{|B|}\sum \mathcal{L}(Y_{\alpha}-Q_{f}^{\alpha}(\cdot,C_{\beta};\theta_{\alpha}, \theta_{f}))
\label{eq.14}
\end{equation}
\begin{equation}
\label{eq.15}
L_{\beta}(\theta_{\beta},\theta_{f})=\frac{1}{|B|}\sum \mathcal{L}(Y_{\alpha}-Q_{f}^{\beta} (\cdot,C_{\alpha};\theta_{\beta}, \theta_{f}))
\end{equation}\par
Where $|B|$ represents the batch size during training, $Y_{\alpha}=r_{\alpha}+\gamma \mathop{max}\limits_{A_{\alpha}'\in \mathcal{A_{\alpha}} }  \sum_{j\in A_{\alpha}'}P(j|s_{\alpha}',A_{\alpha}')Q_{\alpha}^{f} ( s_{\alpha}',j )$ denotes the target Q-value. \textbf{\textbf{In Eq.(\ref{eq.14}) and Eq.(\ref{eq.15}), the updates of $Q_{\alpha}$ and $Q_{\beta}$ both depend on $r_{\alpha}$ since agent $\beta$ has no access to $r_{\beta}$.}}\par

\begin{figure}[!t]
\centering
\includegraphics[width=1.8in]{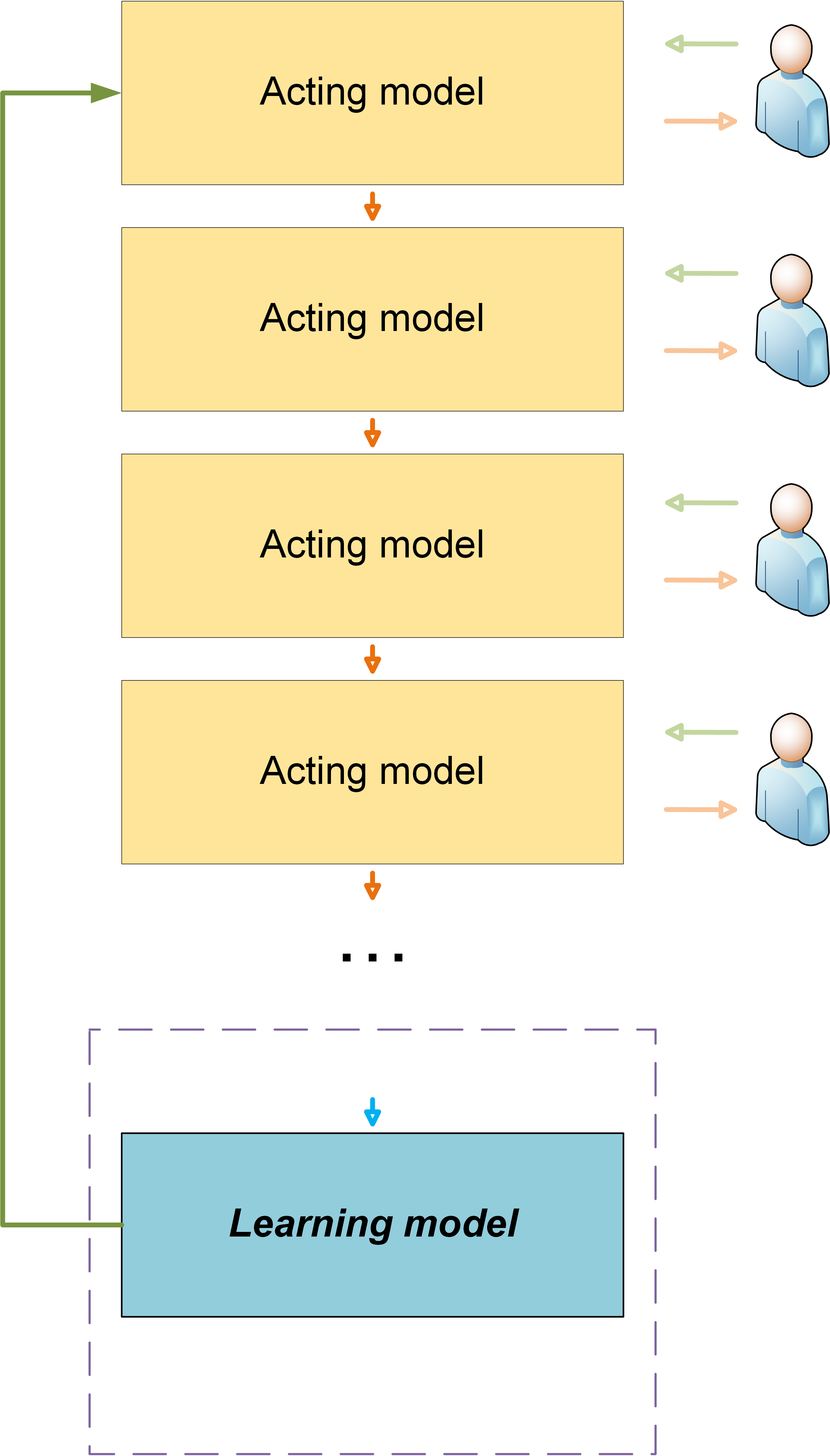}
\caption{\textbf{Main Loop Process of the FedSlate Algorithm.}}
\label{main}
\end{figure}

\paragraph{\textbf{Overview of Acting and Learning}} Our FedSlate algorithm can be divided into two parts: ``acting'' and ``learning'', as shown in Fig.\ref{main}. In the loop of our algorithm, we first perform several rounds of ``acting'' to recommend slates to users and record their feedback. Then, we execute one iteration of ``learning'' to update the network using the collected experiences, thereby optimizing the recommendation strategy. This loop continues for multiple iterations during the training phase. Due to the modular design of our algorithm, we do not strictly differentiate between the training and testing phases. When there is no need for policy optimization, we simply skip the execution of the ``learning'' module in the loop. Similarly, if there are changes in the distribution of input data for the recommendation algorithm (e.g., business adjustments on the platform), the ``learning'' module can be reactivated.\par

\begin{figure}[!t]
\centering
\includegraphics[width=2.8in]{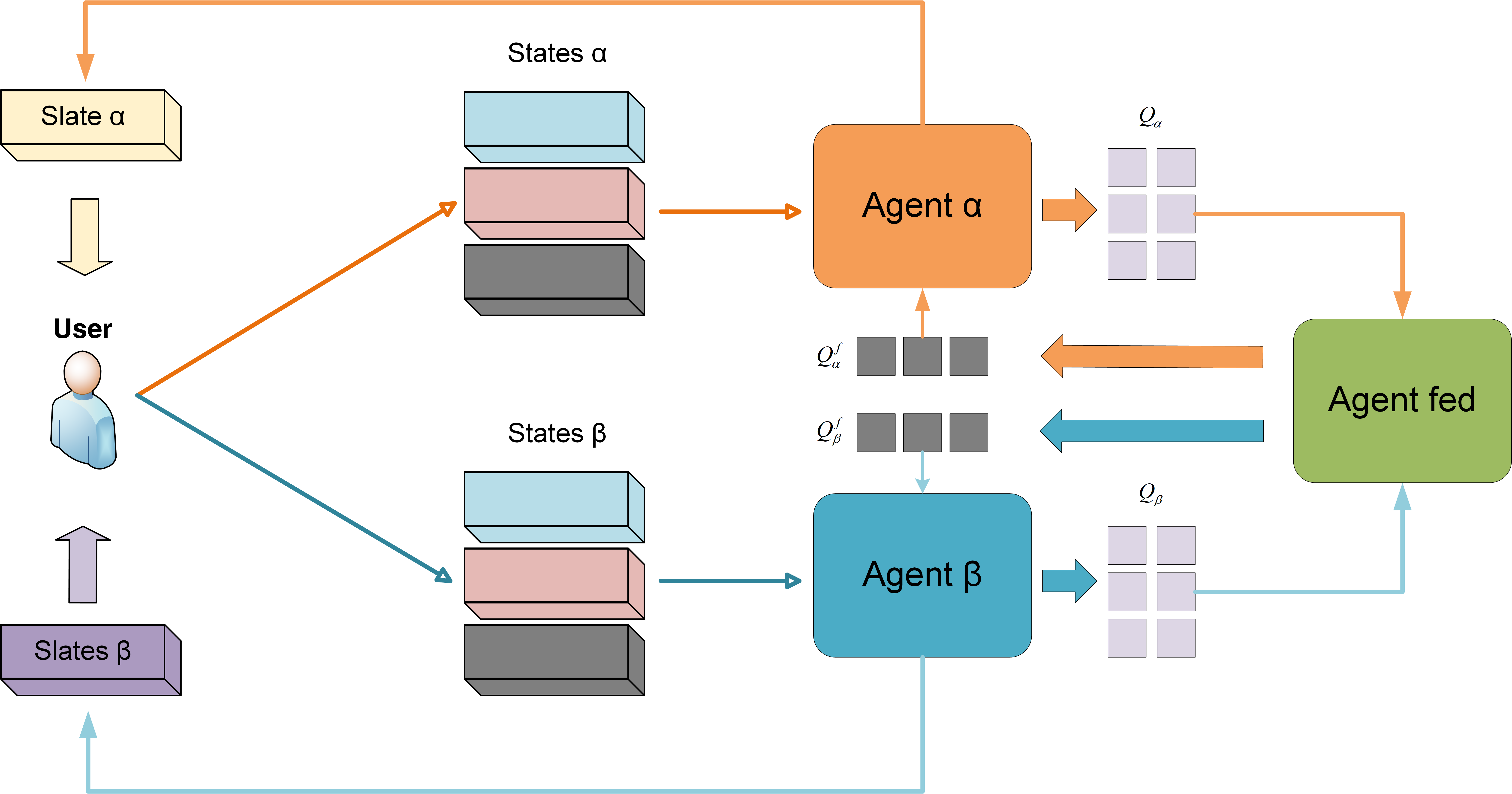}
\caption{\textbf{Overview of the Acting Component in the FedSlate Algorithm.}}

\label{act}
\end{figure}

As shown in Fig.\ref{act},in the ``acting'' phase, agent $fed$ initiates inquiry requests to agent $\alpha$ and $\beta$. Agent $\alpha$ and $\beta$ calculate $Q_{\alpha}$ and $Q_{\beta}$ based on their current states $s_{\alpha}$ and $s_{\beta}$, respectively, and send them to agent $fed$. Agent $fed$ computes $Q_{\alpha}^{f}$ and $Q_{\beta}^{f}$ and sends them back to agents $\alpha$ and $\beta$, respectively. Agent $\alpha$ and $\beta$ construct slates using a greedy method based on the received Q-values and recommend them to the users.\par

\begin{figure}[!t]
\centering
\includegraphics[width=3in]{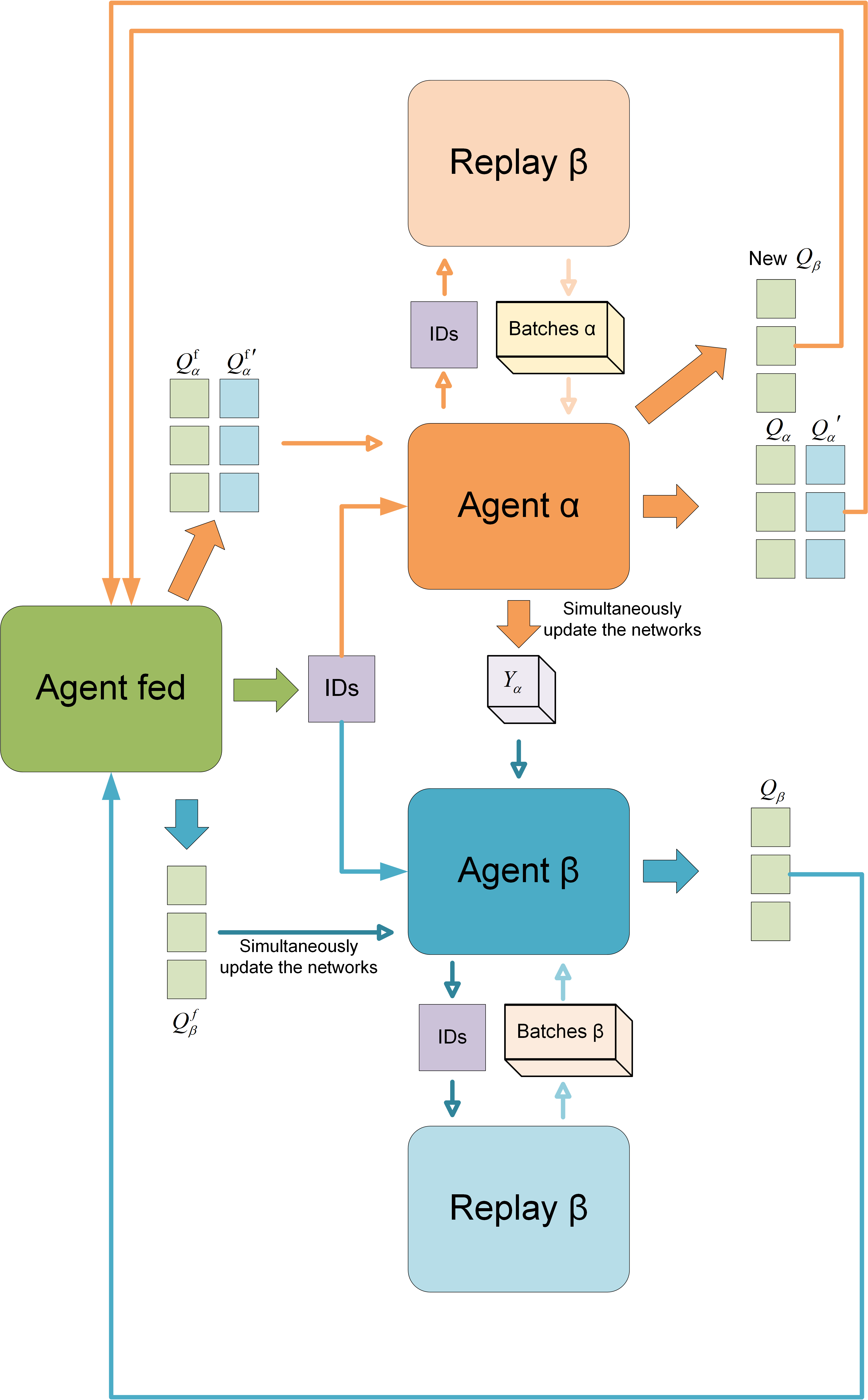}
\caption{\textbf{Overview of the Learning Component in the FedSlate Algorithm.}}

\label{learn}
\end{figure}

During the ``learning'' phase, agents $\alpha$ and $\beta$ are assigned random indices $IDs$ of size $|B|$ by agent $fed$, which correspond to specific training batches ($D_{\alpha}$ and $D_{\beta}$ are in a one-to-one correspondence). Utilizing these batches, agents $\alpha$ and $\beta$ calculate $Q_{\alpha}$, $Q_{\alpha}'$, and $Q_{\beta}$, and subsequently transmit these values to agent $fed$ for the computation of $Q_{f}^{\alpha}$ and ${Q_{f}^{\alpha}}'$. Notably, $Q_{\alpha}'$ and ${Q_{f}^{\alpha}}'$ denote the subsequent Q-values. Agent $fed$ then returns $Q_{f}^{\alpha}$ and ${Q_{f}^{\alpha}}'$ to agent $\alpha$ for the derivation of $Y_{\alpha}$ and the updating of the networks $Q_{\alpha}(s_{\alpha};\theta_{\alpha})$ and $Q_{f}$. After these updates, $Y_{\alpha}$ is conveyed to agent $\beta$, while agent $\alpha$ formulates a refreshed $Q_{\alpha}$ using the newly updated network to forward to agent $fed$. Utilizing the updated $Q_{\alpha}$ and the initial $Q_{\beta}$, agent $fed$ calculates $Q_{\beta}^{f}$ and dispatches it to agent $\beta$, who then updates the networks $Q_{\beta}(s_{\beta};\theta_{\beta})$ and $Q_{f}$ with the aid of $Y_{\alpha}$ and $Q_{\beta}^{f}$. The learning protocol mandates a single update for each local network and two updates for the global network. Figure \ref{learn} provides a schematic representation of the ``learning'' process.

The detailed acting and learning processes can be found in Algorithm \ref{algorithm_1}, \ref{algorithm_2} and \ref{algorithm_3}, where we will introduce some crucial details. Firstly, in step 21 of Algorithm \ref{algorithm_1}, the agent $\alpha$ computes $Y_{\alpha}$ using Eq.(\ref{eq7}),(\ref{eq8}). The inputs $Q_{f}^{\alpha}$ and ${Q_{f}^{\alpha}}'$ received by agent $\alpha$ are both tensors of size $[B,\mathcal{I}]$, where $\mathcal{I}$ represents the size of the candidate documents. Secondly, since we employ the Q-values of items to update our Q-network, a certain transformation must be applied to $Q_{f}^{\alpha}$ and ${Q_{f}^{\alpha}}'$. We only consider the Q-values of items that have been actually selected by the user as the online Q-values (the left-hand side of Eq.(\ref{eq7})), while disregarding the Q-values of items that were not chosen by the user. Thirdly, in order to calculate the target Q-values (the right-hand side of Eq.(\ref{eq7})), we first generate a slate, then compute the probabilities for each recommended item on the slate, and finally take the inner product of these probabilities with their corresponding item's Q-values. This inner product serves as the target Q-value and is utilized in step 21 of Algorithm \ref{algorithm_2} to address the problem of agent $\beta$ being unable to determine the online Q-values for TD updates based on the user's selected item, given the assumption that agent $\beta$ cannot access user feedback.\par

\begin{algorithm}
\caption{FedSlate-ALPHA}\label{algorithm_1}

\begin{algorithmic}[1]
\REQUIRE $S_{\alpha}$
\ENSURE $None$
\STATE \textbf{function} $Init()$
\STATE \hspace{1em} Initialize $Q_{\alpha}$ with random values for $\theta_{\alpha}$
\STATE \textbf{end function}
\STATE \textbf{function} $ComputeQAlpha()$
\STATE \hspace{1em} Observe $s_{\alpha}$
\STATE \hspace{1em} Compute $Q_{\alpha}(s_{\alpha};\theta_{\alpha})$
\STATE \hspace{1em} \textbf{return} $Q_{\alpha}$
\STATE \textbf{end function}
\STATE \textbf{function} $Recommend Slate(Q_{\alpha}^{f})$
\STATE \hspace{1em} Observe $\mathcal{I}_{\alpha}$
\STATE \hspace{1em} Construct Slate $A_{\alpha}$ based on Eq.(\ref{eq8})
\STATE \hspace{1em} Recommend Slate $A_{\alpha}$, obtain state $s_{\alpha}'$ and reward $r_{\alpha}$
\STATE \hspace{1em} Store $(s_{\alpha},A_{\alpha},r_{\alpha},s_{\alpha}')$ in $D_{\alpha}$
\STATE \textbf{end function}
\STATE \textbf{function} $ComputeQAlphaBatch(IDs)$
\STATE \hspace{1em} Sample batch of $D_{\alpha}$ based on indices $IDs$
\STATE \hspace{1em} Compute $Q_{\alpha}(s_{\alpha};\theta_{\alpha})$ and $Q_{\alpha}'(s_{\alpha}';\theta_{\alpha}')$
\STATE \hspace{1em} \textbf{return} batches of $Q_{\alpha}$, $Q_{\alpha}'$
\STATE \textbf{end function}
\STATE \textbf{function} $UpdateQNet(Q_{f}^{\alpha},{Q_{f}^{\alpha}}')$
\STATE \hspace{1em} Compute $Y_{\alpha}$ based on Eq.(\ref{eq7},\ref{eq8})
\STATE \hspace{1em} Update $Q_{\alpha}$ and $Q_{f}$ based on Eq.(\ref{eq7})
\STATE \hspace{1em} \textbf{return} $Y_{\alpha}$
\STATE \textbf{end function}
\end{algorithmic}

\end{algorithm}

\begin{algorithm}
\caption{FedSlate-BETA}\label{algorithm_2}
\begin{algorithmic}[1]
\REQUIRE $S_{\beta}$
\ENSURE $None$
\STATE \textbf{function} $Init()$
\STATE \hspace{1em} Initialize $Q_{\beta}$ with random values for $\theta_{\beta}$
\STATE \textbf{end function}
\STATE \textbf{function} $ComputeQBeta()$
\STATE \hspace{1em} Observe $s_{\beta}$
\STATE \hspace{1em} Compute $Q_{\beta}(s_{\beta};\theta_{\beta})$
\STATE \hspace{1em} \textbf{return} $Q_{\beta}$
\STATE \textbf{end function}
\STATE \textbf{function} $Recommend Slate(Q_{\beta}^{f})$
\STATE \hspace{1em} Observe $\mathcal{I}_{\beta}$
\STATE \hspace{1em} Construct Slate $A_{\beta}$ based on Eq.(\ref{eq8})
\STATE \hspace{1em} Recommend Slate $A_{\beta}$
\STATE \hspace{1em} Store $(s_{\beta},A_{\beta})$ in $D_{\beta}$
\STATE \textbf{end function}
\STATE \textbf{function} $ComputeQBetaBatch(IDs)$
\STATE \hspace{1em} Sample batch of $D_{\beta}$ based on indices $IDs$
\STATE \hspace{1em} Compute $Q_{\beta}(s_{\beta};\theta_{\beta})$
\STATE \hspace{1em} \textbf{return} batches of $Q_{\beta}$
\STATE \textbf{end function}
\STATE \textbf{function} $UpdateQNet(Q_{f}^{\beta},Y_{\alpha})$
\STATE \hspace{1em} Update $Q_{\beta}$ and $Q_{f}$ based on Eq.(\ref{eq7})
\STATE \textbf{end function}
\end{algorithmic}
\end{algorithm}

\begin{algorithm}
\caption{FedSlate-FED}\label{algorithm_3}
\begin{algorithmic}[1]
\REQUIRE{Boolean value to determine whether to cancel the learn module: $is\_learn$,\\
Integer type representing the learning interval: $learn\_every$}
\ENSURE{$None$}
\STATE Initialize $Q_{f}$ with random values for $\theta_{f}$
\STATE Call FedSlate-ALPHA.$Init()$,FedSlate-BETA.$Init()$
\FOR{episode = $1$ to $M$}
\STATE Initialize $step=0$
\WHILE{True}
\STATE Call $Q_{\alpha}=$ FedSlate-ALPHA.$ComputeQAlpha()$
\STATE Call $Q_{\beta}=$ FedSlate-BETA.$ComputeQBeta()$
\STATE Compute $Q_{f}^{\alpha}$, $Q_{f}^{\beta}$ according to Eq.(\ref{eq11},\ref{eq12})
\STATE Call FedSlate-ALPHA.$RecommendSlate(Q_{\alpha}^{f})$
\STATE Call FedSlate-BETA.$RecommendSlate(Q_{\beta}^{f})$
\IF{$is\_learn=True$, $step \bmod learn\_every = 0$}
\STATE Generate $IDs$ randomly
\STATE Call $Q_{\alpha}$, $Q_{\alpha}'$ \\
$=$ FedSlate-ALPHA.$ComputeQAlphaBatch(IDs)$
\STATE Call $Q_{\beta}$ \\
$=$ FedSlate-BETA.$ComputeQBetaBatch(IDs)$
\STATE Compute $Q_{\alpha}^{f}$, ${Q_{\alpha}^{f}}'$ based on Eq.(\ref{eq11})
\STATE Call $Y_{\alpha}$ \\
$=$ FedSlate-ALPHA.$UpdateQNet(Q_{\alpha}^{f}$, ${Q_{\alpha}^{f}}')$
\STATE Call $Q_{\alpha}$ \\
$=$ FedSlate-ALPHA.$ComputeQAlphaBatch(IDs)$
\STATE Compute $Q_{\beta}^{f}$ based on Eq.(\ref{eq12})
\STATE Call FedSlate-BETA.$UpdateQNet(Q_{\beta}^{f},Y_{\alpha})$
\ENDIF
\IF{terminal}
\STATE \textbf{break}
\ENDIF
\STATE Let $step = step+1$
\ENDWHILE
\ENDFOR
\end{algorithmic}
\end{algorithm}

\begin{algorithm}
\caption{FedSlate-FED(extended)}\label{algorithm_4}
\begin{algorithmic}[1]
\REQUIRE{Boolean value to determine whether to cancel the learn module: $is\_learn$,\\ 
Integer type representing the learning interval: $learn\_every$}
\ENSURE{$None$}
\STATE Initialize $Q_{f}$ with random values for $\theta_{f}$
\STATE Call FedSlate-ALPHA.$Init()$,FedSlate-BETA.$Init()$
\FOR{episode=$1$ to $M$}
\STATE Initialize $step=0$
\WHILE{True}
\STATE Call $Q_{\alpha}=$ FedSlate-ALPHA.$ComputeQAlpha()$
\STATE Call $Q_{\beta}=$ FedSlate-BETA.$ComputeQBeta()$
\STATE Compute $Q_{f}^{\alpha}$, $Q_{f}^{\beta}$ according to Eq.(\ref{eq11},\ref{eq12})
\STATE Call FedSlate-ALPHA.$RecommendSlate(Q_{\alpha}^{f})$
\STATE Call FedSlate-BETA.$RecommendSlate(Q_{\beta}^{f})$
\IF{$is\_learn=True$, $step \bmod learn\_every = 0$}
\STATE Generate $IDs$ randomly
\STATE Call $Q_{\alpha}$, $Q_{\alpha}'$ \\
$=$ FedSlate-ALPHA.$ComputeQAlphaBatch(IDs)$
\STATE Call $Q_{\beta}$, $Q_{\beta}'$ \\
$=$  FedSlate-BETA.$ComputeQBetaBatch(IDs)$
\STATE Compute $Q_{\alpha}^{f}$, ${Q_{\alpha}^{f}}'$ based on Eq.(\ref{eq11})
\STATE Call $Y_{\alpha}$ \\
$=$ FedSlate-ALPHA.$UpdateQNet(Q_{\alpha}^{f}$, ${Q_{\alpha}^{f}}')$
\STATE Call $Q_{\alpha}$ \\
$=$ FedSlate-ALPHA.$ComputeQAlphaBatch(IDs)$
\STATE Compute $Q_{\beta}^{f}$, ${Q_{\beta}^{f}}'$ based on Eq.(\ref{eq12})
\STATE Call FedSlate-BETA.$UpdateQNet(Q_{\beta}^{f},{Q_{\beta}^{f}}')$
\ENDIF
\IF{terminal}
\STATE \textbf{break}
\ENDIF
\STATE Let $step = step+1$
\ENDWHILE
\ENDFOR
\end{algorithmic}
\end{algorithm}

In addition, as mentioned by \cite{wang2022individual}, in real-world multi-agent systems, team rewards suffer from sparsity, making it difficult for algorithms to learn a successful team strategy to enhance overall reward. In our setting, if the rewards from Platform A and Platform B are too sparse (in other words, the information received by users on Platform A does not affect their reactions on Platform B), it is not possible to train multiple agents simultaneously using the reward from Platform A. Therefore, our algorithm has a simple variant to address this issue: Algorithm \ref{algorithm_1} is duplicated to replace Algorithm \ref{algorithm_2}, and some minor modifications are made to Algorithm \ref{algorithm_3}. Please refer to Algorithm \ref{algorithm_4} for the specific changes made to Algorithm \ref{algorithm_3}. By simple extension, our algorithm is capable of handling the sparsity issue of team rewards. However, it should be noted that the extended algorithm requires platform B to have access to rewards as well.

Although our method's description focuses on a single user navigating content across multiple platforms, RL-based recommendation algorithms are indeed applicable in industrial-level systems. Such systems generally comprise three key components \cite{RN1}: a user feedback log system; offline DNN training using these logs (typically for scoring user-item pairs); and online recommendation services that perform real-time scoring and ranking to select the top-k items. FedSlate, a Q-learning based algorithm optimized for long-term benefits, constructs its state space with features that predict immediate user feedback rewards and are capable of self-prediction. These characteristics allow for the reuse of existing log systems. Addressing user generalization, FedSlate adapts the MDP for all users, leveraging the generalization capabilities already present in existing systems. Regarding user feedback modeling, FedSlate utilizes existing ranking system definitions, such as predicted Click-Through Rate (pCTR), allowing the system to employ these conventional models to predict the likelihood of an item being selected.

\section{Experimental Setup}
\label{Experimental Setup}
In this section, the RecSim platform is utilized to construct a simulation environment aimed at evaluating the efficacy of our FedSlate algorithm. Specifically, detailed information regarding the simulation environment is provided. Subsequently, a comparative analysis is conducted between our algorithm and the SlateQ method, to ascertain whether agent $\alpha$ demonstrates enhanced performance subsequent to its participation in the FL process, as compared to its individual learning performance. A metric is proposed to assess this particular aspect. Furthermore, a comparison is made between FedSlate and a purely random recommendation approach, in order to ascertain if agent $\beta$ (as agent $\beta$ lacks access to user feedback and therefore cannot optimize the recommendation strategy using conventional methods) can derive benefits from the federated setting. By considering these two aspects, we aim to demonstrate whether our algorithm can effectively exploit the consistency of user behavior across different platforms. Lastly, to address the potential sparsity issue of team rewards, the performance of our extended FedSlate algorithm is evaluated within a more complex scenario and compared against the performance of the basic FedSlate algorithm under conditions of sparse team rewards.

\subsection{Simulation Environment}
RecSim\cite{RN26} is a simulation platform for constructing and evaluating recommendation systems that naturally support sequential interactions with users. Developed by Google, it simulates users and environments to assess the effectiveness and performance of recommendation algorithms. We employ RecSim to create an environment that reflects user behavior and item structure to evaluate our FedSlate algorithm.\par

We construct a ``Choc vs. Kale'' recommendation scenario, where the goal is to maximize user satisfaction and engagement over the long term by recommending a certain proportion of ``chocolate'' and ``kale'' elements. In this scenario, the ``chocolate'' element represents content that is interesting but not conducive to long-term satisfaction, while the ``kale'' element represents relatively less exciting but beneficial content for long-term satisfaction. The recommendation algorithm needs to balance these two elements to achieve maximized long-term user satisfaction. We believe this scenario aligns well with our assumption that ``user responses to content on other platforms are influenced to some extent by the content they are exposed to on the current platform''. If a platform consistently recommends ``chocolate'' content to users, their long-term satisfaction is likely to be compromised.\par

In our scenario, the entire simulation environment consists primarily of document models and user models. The document model serves as the main interface for interaction between users and the recommendation system (agent) and is responsible for selecting a subset of documents from a database containing a large number of documents to deliver to the recommendation system. The user model simulates user behavior and reacts to the slates provided by the recommendation system.\par

The database in the document model essentially serves as a container for observable and unobservable features of underlying documents. In our scenario, document attributes are modeled as continuous features with values in the range of $[0,1]$, referred to as the Kaleness scale. A document with a score of 0 represents pure ``chocolate'', which is interesting but regretful, while a document with a score of 1 represents pure ``kale'', which is less exciting but nutritious. Additionally, each document has a unique integer ID, and the document model selects $N$ candidate documents in sequential order based on their IDs.\par

The user model includes unobservable and observable features of users. Based on these features, the model responds to the received slate according to certain rules. Each user is characterized by the features of net kale exposure ($nke_{t}$) and satisfaction ($sat_{t}$), which are associated through the sigmoid function $\sigma$ to ensure that $sat_{t}$ is constrained within a bounded range. Specifically, the satisfaction level is modeled as a sigmoid function of the net kale exposure, which determines the user's satisfaction with the recommended slate:\par

\begin{equation}
    sat_{t} = \sigma (\tau \cdot nke_{t})
\end{equation}\par

Where, $\tau$ is a user-specific sensitivity parameter. Upon receiving a Slate from the recommendation system, users select items to consume based on the Kaleness scale of the documents. Specifically, for item $i$, the probability of it being chosen is determined by $p \sim e^{1-kaleness(i)}$. After making their selections, the net kale exposure evolves as follows:

\begin{equation}
    nke_{t+1} = \beta \cdot nke_{t} + 2(k_{i}-1/2) + \mathcal{N}(0,\eta)
\end{equation}\par

Where, $\beta$ represents a user-specific memory discount, while $k_{i}$ corresponds to the kaleness of the selected item, and $\eta$ denotes some noise standard deviation. Lastly, our focus will be on the user's engagement $s_{i}$, i.e. a log-normal distribution with parameters linearly interpolating between the pure kale response $(\mu_{k}, \sigma_{k})$ and the pure choc response $(\mu_{c},\sigma_{c})$:\par
\begin{equation}
    s_{i} \sim log\mathcal{N}(k_{i}\mu_{k}+(1-k_{i})\mu_{c},k_{i}\sigma_{k}+(1-k_{i})\sigma_{c})
\end{equation}
The satisfaction variable $sat_{t}$ represents the sole dynamic component of the user's state, and thus, we generate the user's observable state based on it. In the simulation, user satisfaction is modeled and computed as a latent state. However, to simulate real-world scenarios, we map the latent state to an observable state by introducing noise to account for user uncertainty.\par

We will develop two distinct document models, representing Platform A and Platform B, respectively, and integrate them with a user model to establish a comprehensive environment. Furthermore, we will introduce a time budget parameter to the user model, constraining the browsing duration for users. Upon a user's cessation of browsing, the environment will generate a terminal signal to indicate the completion of the interaction.\par

The evaluation environment for the baseline method (i.e., SlateQ) is intentionally simplified compared to the previously described environments. It comprises a single document model and a single user model, allowing for the simulation of sequential interaction behavior of an individual user on a solitary platform.

\subsection{Algorithm Evaluation}
\label{exp:main}
\begin{figure}[htbp]
\centering
\includegraphics[width=3.4in]{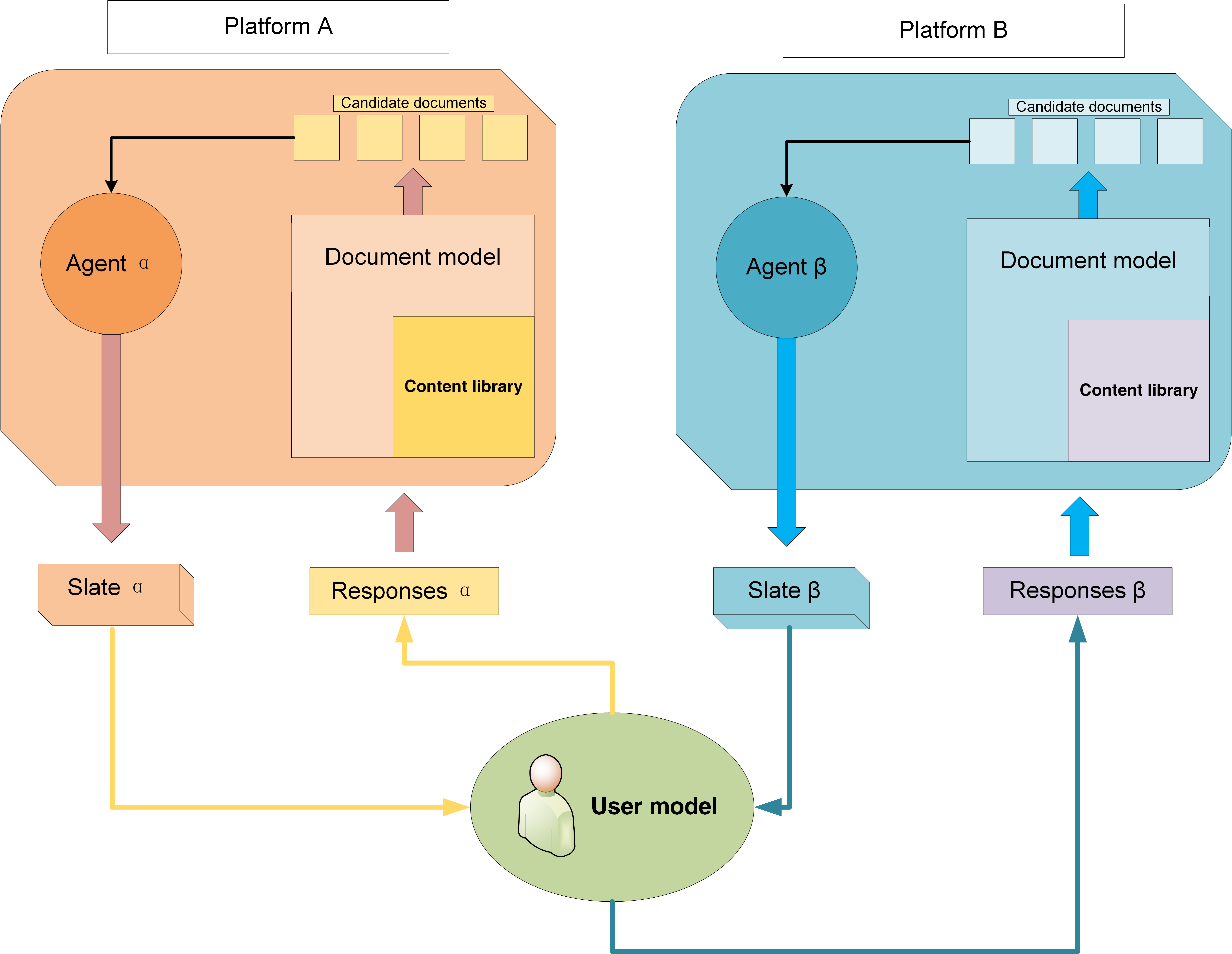}
\caption{\textbf{Interaction Process between Environment and Agent.}}
\label{env}
\end{figure}

We evaluate our algorithm in a simulated environment,the schematic representation of the interactive process between the environment and the agent is depicted as illustrated in Fig.\ref{env}. First, let's define the states, actions, and rewards.

\begin{itemize}
\item \textbf{States:} The environment state observed by Agent $\alpha$ consists of the user's observable state (with a count of $1$) and the features of candidate documents (with a count of $N$). Since Platform A has access to user feedback information, we incorporate the user's historical engagement records into the observable state. Specifically, we include the user's previous 5 engages in the state. Therefore, the state $\alpha$ corresponding to Platform A is a tensor of size $[1+5 \times n + N]$, where $n$ represents the slate size we set. Platform B, which lacks access to user feedback information, does not include the user's historical engagement records in the observed environment state. Hence, the state $\beta$ corresponding to Platform B is a tensor of size $[1 + N]$, containing only the current user's observable state and the features of candidate documents.

\item \textbf{Actions:} The actions for both Agent $\alpha$ and Agent $\beta$ are similar. They involve recommending a slate of content determined by the algorithm. The actions are essentially an integer tensor of size $[n]$, representing the specific item IDs that form the slate.

\item \textbf{Rewards:} As our objective is to maximize long-term user satisfaction, we employ cumulative user engagement as the reward. It is important to note that the environment provides both user engages on Platform A and Platform B simultaneously. However, for Agent $\beta$, the engages on Platform B are completely inaccessible. We include this information in the environment solely for evaluating the performance of our algorithm and not for training our model.

\item \textbf{Criteria (Agent $\alpha$):} To assess the learning time consumption of Agent $\alpha$ in our algorithm, we introduce a metric called ``Episodes to Reach Optimal Reward'' (ETROR), denoted as $M^{'}$. Let $M_{1}$ represent the number of episodes undergone by the baseline method (SlateQ) during training, and $M_{2}$ represent the number of episodes undergone by FedSlate during training. Similarly, we define $R_{M_{1}}$ and $R_{M_{2}}$ as the rewards achieved at time $M$ during the training process. For the baseline method, if $R_{M_{1}} \ge R_{M_{1}+t},t \in \mathbb{N}^{+}$, we consider $M_{1}=M_{1}^{'}$. Taking into account the instability of reinforcement learning, we add a small positive integer term $\varepsilon $ to the inequality, i.e., $R_{M_{1}}+\varepsilon \ge R_{M_{1}+t},t \in \mathbb{N}^{+}$. In such cases, we consider $M_{1}$ as the number of episodes consumed by the baseline method to reach optimal rewards. Similar definitions can be applied to determine $M_{2}^{'}$.

\item \textbf{Criteria (Agent $\beta$):} We denote $R_{M_{3}^{'}}$ as the optimal reward achieved by Agent $\beta$ with the introduction of FL at $M_{3}^{'}$. We define $R_{rnd}$ as the average rewards obtained by Platform B using a random recommendation method over episodes $[0, M_{3}^{'}]$. If $R_{M_{3}^{'}} \ge R_{rnd}$, we conclude that Agent $\beta$ benefits from the FL.

\end{itemize}

\begin{figure*}[!t]
\centering
\includegraphics[width=7in]{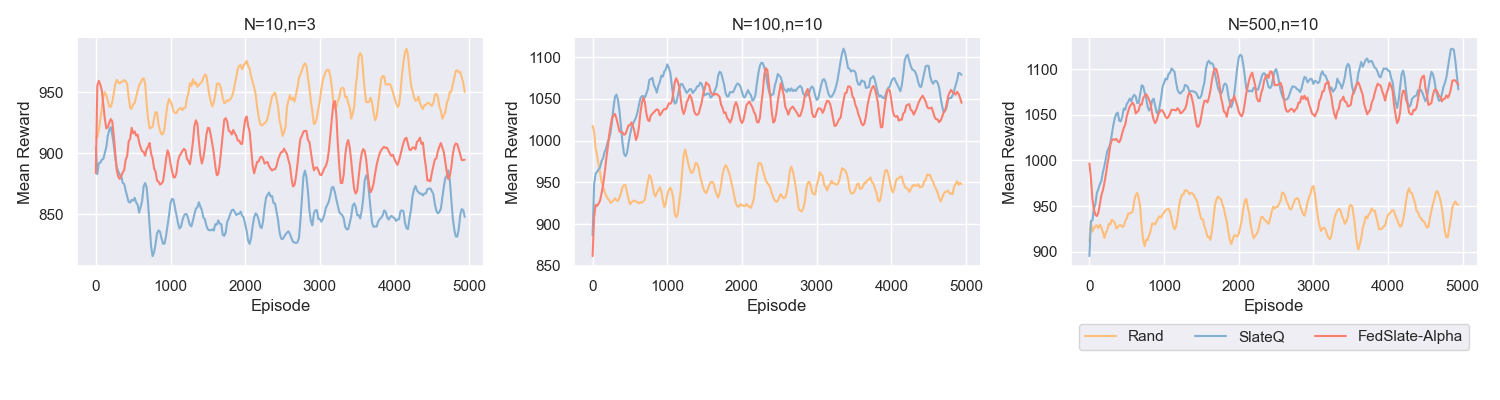}
\caption{\textbf{Evaluation of FedSlate versus Baseline Performance under Various Environmental Conditions.}}
\label{cop1}
\end{figure*}

\textbf{Experimental Results:} We conducted multiple iterations of FedSlate and a baseline method under various environmental settings. \textit{Specifically, for the parameter selection of $N$ and $n$, we adhered to the standard configurations detailed in the RecSim technical documentation\cite{RN26}.} Our comparison focused on the performance metrics $M_{1}^{'}$ and $M_{2}^{'}$. As depicted in Table \ref{table1}, $M_{2}^{'}$ is consistently less than or equal to $M_{1}^{'}$ across different settings, indicating that FedSlate can enhance the training velocity of agent $\alpha$. However, it may compromise the agent's learning of an optimal local policy by potentially reducing the optimal reward. This is attributed to FedSlate's design, which focuses on optimizing the aggregate long-term benefit for users across platforms rather than maximizing the immediate value for individual users on a single platform. Notably, we omitted the performance comparison between the SlateQ and FedSlate algorithms in the scenario with $N=10,n=3$, since neither algorithm converged in this setting, performing substantially worse than a random recommendation approach. This lack of convergence is likely due to overfitting within an overly simplistic environment. Despite this, We argue that FedSlate effectively aids feedback data owners—designated as agent $\alpha$ in the experiment—by improving training efficiency and significantly reducing computational resource consumption. Moreover, the recommendation strategy learned through FedSlate demonstrates comparable performance to that developed using SlateQ. For a visual representation of the comparative experimental results under different settings, please refer to Fig. \ref{cop1}. \textbf{In the context of random recommendation algorithms, it is noteworthy that we compare their average reward against the optimal reward achieved by FedSlate and SlateQ, while refraining from providing its ETROR.} This approach is inherently justified, given that for random recommendation algorithms, there is no process of `learning an optimal recommendation strategy'. Instead, they persistently operate with suboptimal performance, making the utilization of average reward a more representative metric for evaluating the performance of random algorithms.
\par

\begin{table}[htbp]
	\centering
	\caption{Comparison of Performance between FedSlate and Baseline under Various Environmental Settings}
\begin{adjustbox}{width=0.45\textwidth}
\begin{tabular}{lcrrr}
\toprule
\multirow{2}{*}{\textbf{Metric}} & \multirow{2}{*}{\textbf{Method}} & \multicolumn{3}{c}{\textbf{Environment}} \\
\cmidrule(lr){3-5}
& & {N=10,n=3} & {N=100,n=10} & {N=500,n=10} \\
\midrule
\multirow{2}{*}{ETROR} & SlateQ\cite{RN1} & N/A & 3400 & 2020 \\
& FedSlate & N/A & \textbf{2340} & \textbf{1700} \\
\midrule
\multirow{2}{*}{Optimal Reward} & SlateQ & N/A & \textbf{1115.072} & \textbf{1117.974} \\
& FedSlate & N/A & 1106.57 & 1107.092 \\
\midrule
Mean Reward & Rand & 947.087 & 942.75 & 938.665 \\
\bottomrule
\end{tabular}
\end{adjustbox}
\label{table1}   
\end{table}

\begin{figure*}[!t]
\centering
\includegraphics[width=6in]{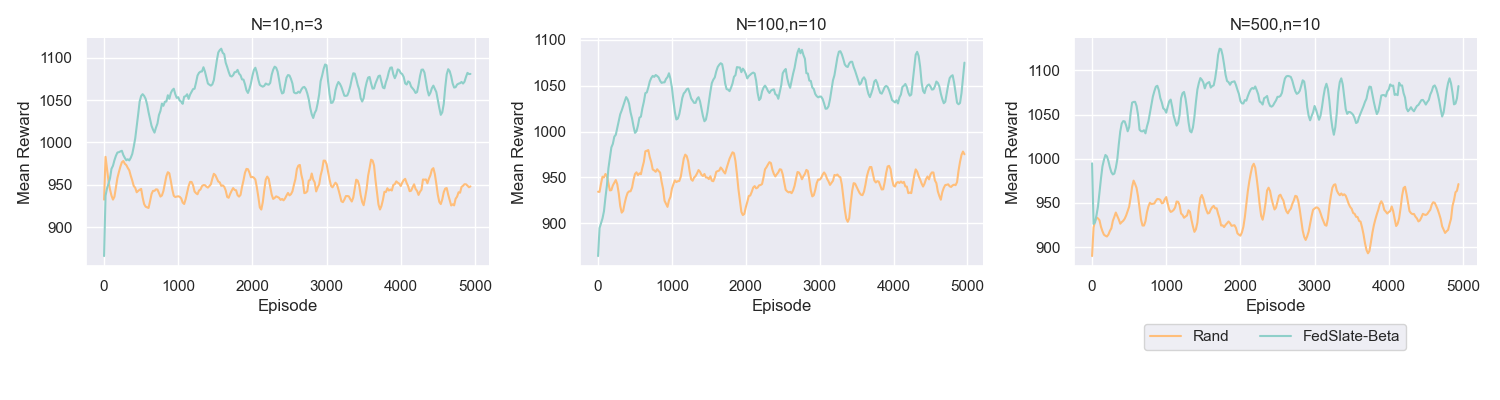}
\caption{\textbf{Evaluation of FedSlate versus Random Recommendation Algorithm Performance under Various Environmental Conditions.}}
\label{cop2}
\end{figure*}

Furthermore, a pivotal aspect of our study is assessing if FedSlate can empower platforms lacking user feedback to augment their recommendation systems by leveraging feedback from other platforms. The performance of agent $\beta$, which employs FedSlate, is contrasted with that of random recommendations. As illustrated in Table \ref{table2}, agent $\beta$ consistently outperforms random recommendations in various settings, as indicated by $R_{M_{3}^{'}} \ge R_{rnd}$. \textbf{Please note that our comparative analysis was strictly limited to evaluating the performance of FedSlate-Beta against the random recommendation algorithm. This limitation was necessitated by a fundamental constraint of the original SlateQ framework, which is its incapacity to develop an effective recommendation policy without reward feedback.} This finding demonstrates that FedSlate enables entities (represented by agent $\beta$) without feedback data to benefit from user feedback. Such data-deprived entities are likely to have a heightened interest in FedSlate, as it offers a means to exploit insights from feedback data previously inaccessible to them. The comparative results are visually represented in Fig. \ref{cop2}. Notably, in scenarios with $N=10,n=3$, agent $\beta$ demonstrates a reliable learning of the recommendation strategy. This robustness can be attributed to the indirect training process where the target Q-values from agent $\alpha$ differ from agent $\beta$'s objective Q-values, adding a layer of complexity to the training.\par

Through comparative experiments, we demonstrate the effectiveness of FedSlate for all participating parties in the FL process, indicating that FedSlate can enhance the performance of recommendation systems by leveraging the correlation of user behaviors across different platforms, without compromising user privacy.\par

\begin{table}[htbp]
	\centering
	\caption{Comparison of Performance between FedSlate and Random Recommendation Algorithm under Various Environmental Settings}
\begin{adjustbox}{width=0.45\textwidth}
\begin{tabular}{lcrrr}
\toprule
& & \multicolumn{3}{c}{\textbf{Environment}} \\
\cmidrule(lr){3-5}
\multirow{-2}{*}{\textbf{Metric}} & \multirow{-2}{*}{\textbf{Method}} & \textbf{N=10,n=3} & \textbf{N=100,n=10} & \textbf{N=500,n=10} \\
\midrule
Optimal Reward & FedSlate & \textbf{1115.281} & \textbf{1102.269} & \textbf{1129.96} \\
Mean Reward & Rand & {948.294} & 944.982 & 939.979 \\
\bottomrule
\end{tabular}
\end{adjustbox}
\label{table2}   
\end{table}

\begin{figure*}
	\centering
	\subfloat[\footnotesize Agent Alpha]{	\includegraphics[width=0.25\textwidth]{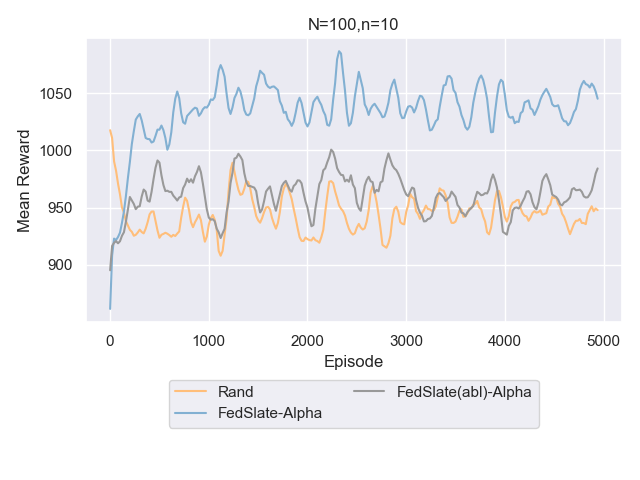}}\hspace{0.8in} 
	\subfloat[\footnotesize Agent Beta]{	\includegraphics[width=0.25\textwidth]{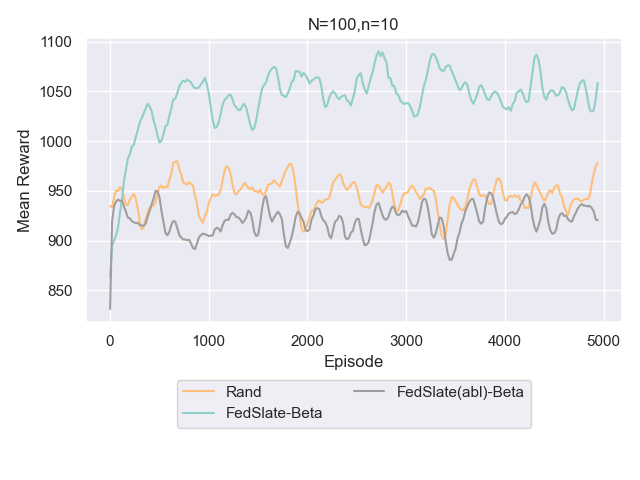}}
	\caption{\textbf{Comparison of Performance between Ablated FedSlate Algorithm and Original Algorithm.}}
	\label{figabl}
\end{figure*}

\begin{figure*}
	\centering
	\subfloat[\footnotesize Agent Alpha]{	\includegraphics[width=0.3\textwidth]{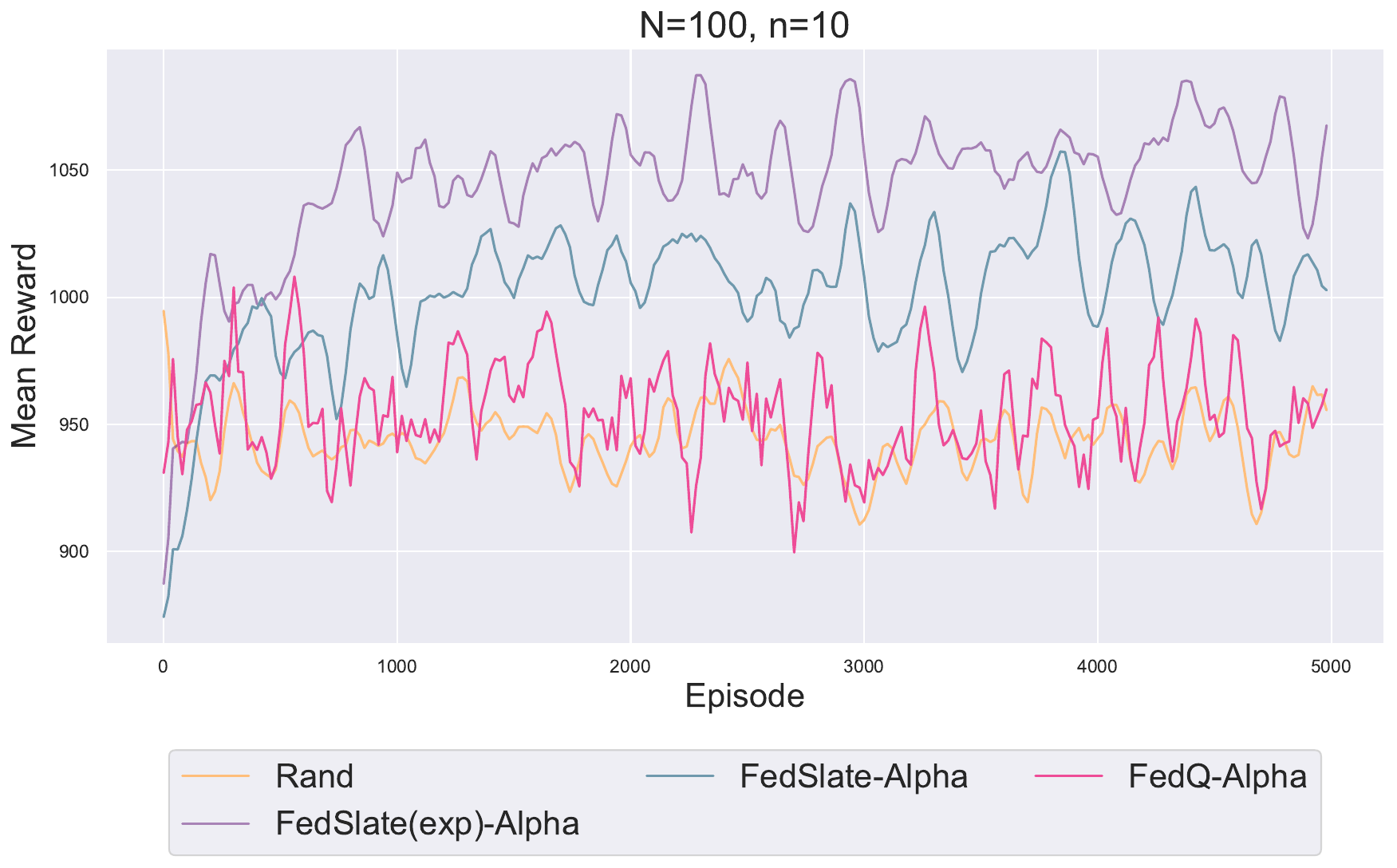}}\hspace{0.8in} 
	\subfloat[\footnotesize Agent Beta]{	\includegraphics[width=0.3\textwidth]{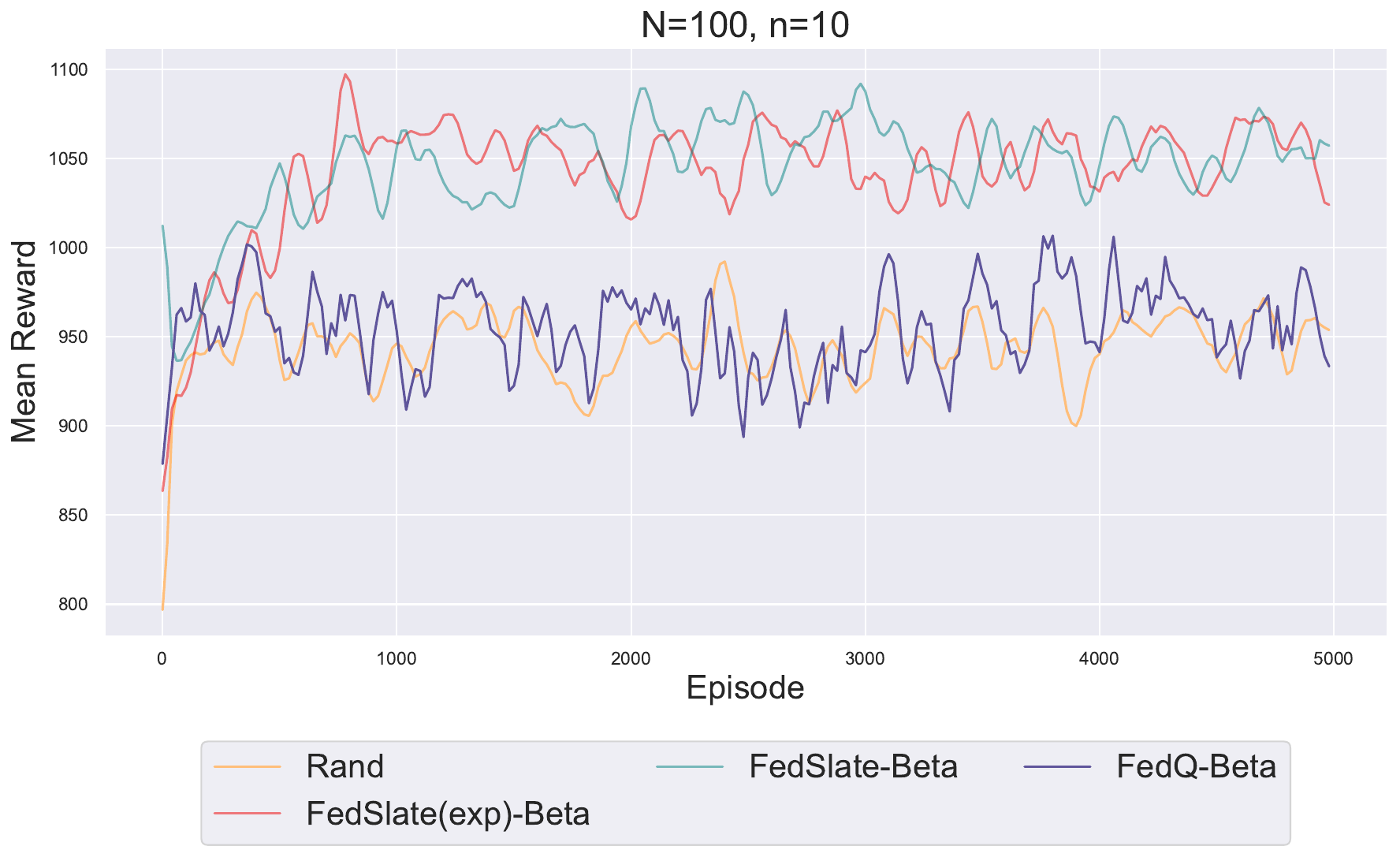}}
	\caption{\textbf{Comparison of Performance between Extended FedSlate Algorithm and Baseline Algorithm.}}
	\label{figexp}
\end{figure*}

\begin{table}[htbp]
	\centering
	\caption{Comparison of Performance between Ablated FedSlate Algorithm and Original Algorithm}
\begin{tabular}{lcr}
\toprule
\multirow{2}{*}{\textbf{Metric}} & \multirow{2}{*}{\textbf{Method}} & \textbf{Environment} \\
\cmidrule(l){3-3}
& & \textbf{N=100,n=10} \\
\midrule
\multirow{4}{*}{ETROR} & FedSlate-Alpha & \textbf{2340} \\
& FedSlate(abl)-Alpha & N/A \\
& FedSlate-Beta & \textbf{2800} \\
& FedSlate(abl)-Beta & N/A \\
\midrule
\multirow{4}{*}{Optimal Reward} & FedSlate-Alpha & \textbf{1115.072} \\
& FedSlate(abl)-Alpha & 1005.524 \\
& FedSlate-Beta & \textbf{1102.269} \\
& FedSlate(abl)-Beta & 956.612 \\
\bottomrule
\end{tabular}
\label{table_abl}
\end{table}

\begin{table}[htbp]
	\centering
	\caption{Comparison of Performance between Extended FedSlate Algorithm and Original Algorithm}
\begin{adjustbox}{width=0.45\textwidth}
\begin{tabular}{llrr}
\toprule
\multirow{2}{*}{\textbf{Metric}} & \multirow{2}{*}{\textbf{Method}} & \multicolumn{2}{c}{\textbf{Environment(N=100,n=10)}} \\
\cmidrule(l){3-4}
& & \textbf{dense} & \textbf{sparse} \\
\midrule
\multirow{4}{*}{ETROR} & FedSlate-Alpha & 2340 & 3540 \\
& FedSlate(exp)-Alpha & N/A & \textbf{2280} \\
& FedSlate-Beta & 2800 & 2940 \\
& FedSlate(exp)-Beta & N/A & \textbf{780} \\
\midrule
\multirow{4}{*}{Optimal Reward} & FedSlate-Alpha & 1115.072 & 1081.422 \\
& FedSlate(exp)-Alpha & N/A & \textbf{1091.515} \\
& FedSlate-Beta & 1102.269 & 1042.926 \\
& FedSlate(exp)-Beta & N/A & \textbf{1108.855} \\
\midrule
Mean Reward & FedQ\cite{DBLP:journals/corr/abs-2303-04689} & 981.971 & 973.150 \\
\bottomrule
\end{tabular}
\end{adjustbox}
\label{table3}
\end{table}

\subsection{Ablation Experiment}

The FedSlate algorithm employs the extraction of information from the outputs of local networks and its transfer to the global network, which then calculates Q-values used for selecting recommended content. Given the operational mechanism of the algorithm, a concern arises regarding whether the global network tends to discard Q-values that do not originate from itself and solely outputs its own Q-values. Should such a situation occur, it would indicate that our FedSlate algorithm, when making recommendations, solely relies on information from individual platforms, akin to performing SlateQ separately on each platform, without engaging in FL. To investigate this matter, we conduct an ablation experiment.

In the evaluation experiment of FedSlate, both the local and global networks are fully connected networks with five hidden layers. We employ the Mish function \cite{misra2019mish} as the activation function Nonetheless, the hidden layers vary in size, with the local network  possessing a larger size and the global network  a smaller size. In the ablation experiment, we maintain the structure of the local network unchanged and simplify the global network as much as possible. We reduce the global network to a single hidden layer network and remove the activation function, transforming it into a simple Linear Model. Through this setup, we force FedSlate to directly utilize Q-values generated by the local network for recommendations. The experimental results, as depicted in Table \ref{table_abl}, do not provide the ETROR data for the ablated algorithm. Throughout the entire training process, the rewards for the ablation group diverge completely, and its performance does not exhibit significant improvement compared to random recommendation algorithms. The detailed experimental results can be referred to in Fig.\ref{figabl}. The ablation experiment demonstrates the effectiveness of our FedSlate framework in terms of vertical FL.

\subsection{Evaluation of Extended Algorithms}
The proposed scenario requires collaborative efforts among multiple agents to optimize the collective LTV of individual users across different platforms. In our implementation, we used Platform A's rewards as a proxy for overall lifetime value. However, sparse team rewards can potentially degrade FedSlate's performance. To investigate this effect, we simulated sparse team reward scenarios by implementing different random seeds for sub-environments within the main environment, conducting experiments with parameters $N=100$ and $n=10$.

Our experimental framework encompasses both the original FedSlate algorithm and an extended version designed to address sparse team rewards. While a comprehensive comparison with state-of-the-art methods would ideally position our approach within the current technological landscape, the integration of FL into recommendation systems remains an emerging field. Our literature review \cite{yang2020federated, DBLP:journals/tce/JaveedSKJII24, DBLP:journals/ojcomps/ChronisVHSANBD24} identified the FedQ \cite{DBLP:journals/corr/abs-2303-04689} algorithm, which leverages global recommendation system insights, as the most relevant comparable work. However, FedQ primarily focuses on applying FL to traditional recommendation models without addressing cross-platform LTV considerations. Due to FedQ's requirement for complete user feedback, we constructed a controlled environment with universal access to user feedback for comparative analysis with our extended FedSlate version.

Table \ref{table3} presents the experimental results. In environments with sparse team rewards, FedSlate agents required extended learning periods to develop recommendation policies, with diminished policy effectiveness. The extended FedSlate algorithm demonstrated superior performance compared to the original version in both ETROR metrics and Optimal Reward calculations, with agent $\beta$ showing particularly marked improvements. Fig. \ref{figexp} illustrates the comparative experimental outcomes. A significant limitation of the FedQ algorithm lies in its inability to account for cross-platform LTV, leading it to optimize for immediate rewards during training. This approach often results in recommending short-term engaging but potentially harmful items (metaphorically termed `chocolate' items), ultimately reducing long-term user engagement and necessitating policy relearning. In the ``Choc vs. Kale'' scenario, FedQ's performance barely exceeded random recommendation benchmarks and exhibited significant instability, underscoring FedSlate's distinctive advantages among current methodologies.

\subsection{Extended Experiments Across Multiple Platforms}

\begin{figure*}
	\centering
	\subfloat[\footnotesize Configuration 1]{	\includegraphics[width=0.35\textwidth]{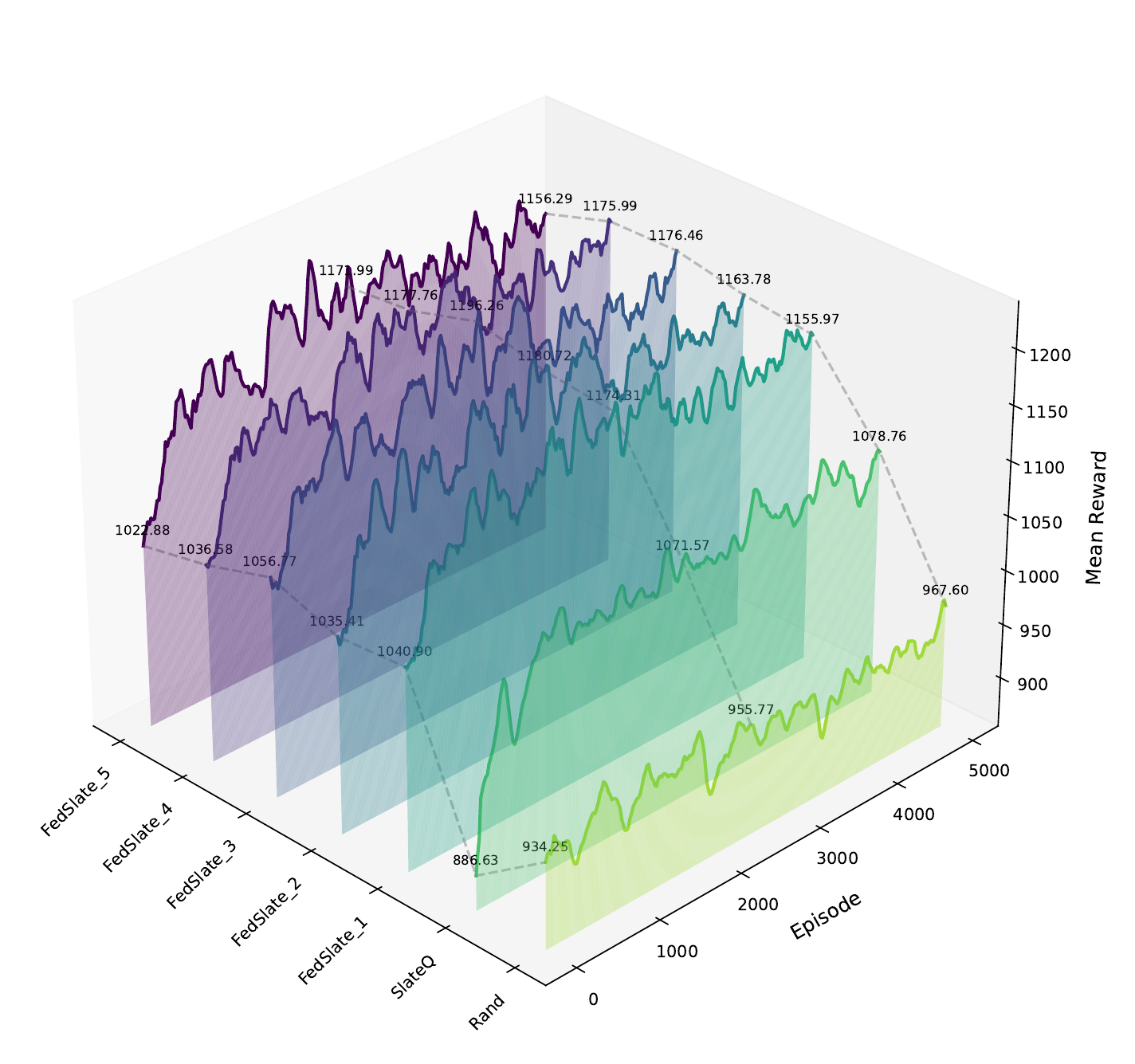}}\hspace{0.8in} 
	\subfloat[\footnotesize Configuration 2]{	\includegraphics[width=0.35\textwidth]{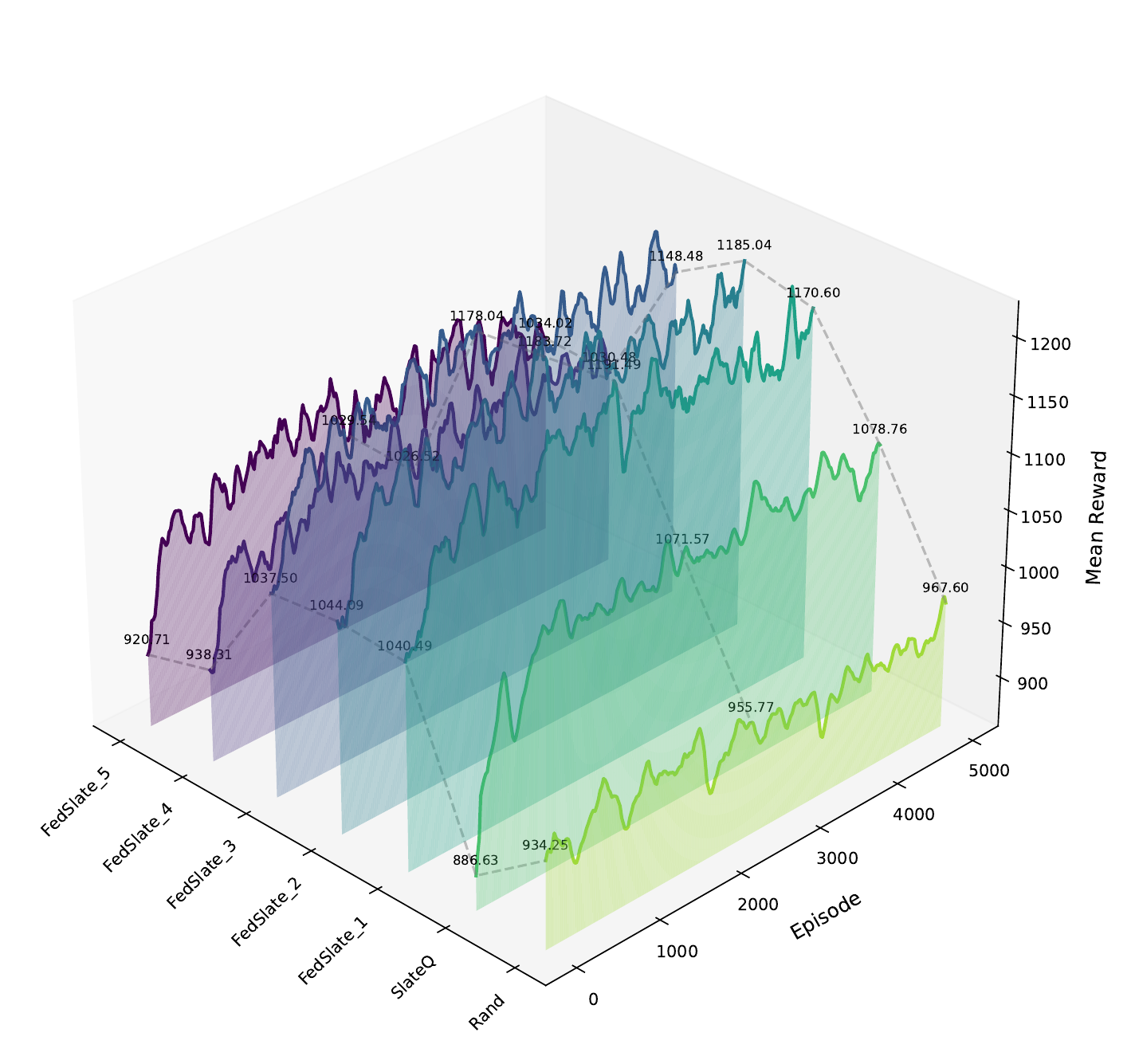}}
	\caption{\textbf{Performance evaluation of FedSlate algorithm across expanded platform configurations.} The designation `FedSlate-1' represents Platform 1's implementation within the federated learning framework. Configuration 1 (5 platforms with user feedback access) exhibited enhanced convergence and robustness in recommendation policy learning relative to baseline metrics. Configuration 2 demonstrated that platforms without direct feedback access (Platforms 4 and 5) achieved performance parity with the original SlateQ algorithm through federated knowledge transfer from Platforms 1-3, establishing the framework's scalability and sustained effectiveness with increased federation size.}
	\label{fig:multplat}
\end{figure*}

To comprehensively evaluate the proposed FedSlate framework, we conducted extended experiments under two distinct configurations. In the first configuration, five platforms with complete access to their respective user feedback participated in the study. This setup employed the extended FedSlate algorithm described in Section \ref{method:fedslate} to investigate potential additional benefits derived from cross-platform feedback sharing. The second configuration involved five platforms, with two platforms lacking access to user feedback, enabling us to assess the framework's scalability and effectiveness with increased participation.

The experimental parameters, including hyperparameters, remained consistent with those detailed in Section \ref{exp:main}, with results presented in Figure \ref{fig:multplat}. Analysis of the first configuration revealed that all participating platforms demonstrated accelerated and more robust learning of recommendation strategies. Furthermore, leveraging cross-platform user LTV metrics resulted in consistently superior recommendation performance compared to baseline methods, as evidenced by increased reward metrics.

In the second configuration, platforms 4 and 5, which lacked direct access to user feedback, achieved performance levels comparable to the original SlateQ algorithm through indirect benefits from platforms 1, 2, and 3. This outcome demonstrates the framework's sustained effectiveness with increased participant numbers. While Section \ref{exp:main} noted that FedSlate might impact the optimal local strategies of feedback-enabled agents despite improving their training efficiency, the results from the second configuration suggest this limitation diminishes with increased federation participation. This improvement can be attributed to the synergistic effects of collaborative user interest development, where the fundamental components of user engagement yield greater benefits as the number of participating platforms increases, thereby enhancing the overall effectiveness of content recommendations.

\section{Conclusion}
To tackle the complexity of integrating user privacy data across diverse platforms into recommendation systems, we introduce a novel reinforcement learning algorithm, designated as FedSlate. This algorithm is designed to develop superior recommendation tactics collaboratively across multiple agents while safeguarding user privacy. Utilizing RecSim, we established a multi-platform recommendation simulation to assess how our algorithm benefits various participants within FL. Our research indicates that FedSlate effectively resolves the challenges of cross-platform learning in recommendation systems without necessitating the exchange of private data between platforms.

\section{Limitations}
Although our FedSlate benefits all participating entities in federated recommendations, the introduction of FL incurs higher communication costs, potentially negating some of the performance enhancements in recommendation systems. Future research will investigate the thresholds for cost-benefit trade-offs. Additionally, the gains from FedSlate are unevenly distributed among the participants in the FL process. Entities lacking access to user feedback benefit disproportionately in terms of recommendation accuracy, receiving greater advantages. Subsequent efforts will focus on promoting fairness within the FL process by evaluating the contributions of each participant. Looking ahead, our objective is to refine and broaden FedSlate's application in privacy-centric cross-platform recommendation systems. To this end, we aim to fortify FedSlate with advanced privacy-preserving technologies such as secure aggregation, differential privacy, and encrypted data in FL, thereby safeguarding user privacy throughout the collaborative learning process, even when handling sensitive data. Additionally, we will undertake extensive testing and evaluation of the FedSlate algorithm in real-world recommendation scenarios across diverse platforms and user demographics, yielding crucial insights into the algorithm's efficiency, scalability, practical viability, and its effects on user satisfaction and engagement.


\bibliographystyle{IEEEtran}
\bibliography{fedslated_ref}

\begin{thebibliography}{10}
\providecommand{\url}[1]{#1}
\csname url@samestyle\endcsname
\providecommand{\newblock}{\relax}
\providecommand{\bibinfo}[2]{#2}
\providecommand{\BIBentrySTDinterwordspacing}{\spaceskip=0pt\relax}
\providecommand{\BIBentryALTinterwordstretchfactor}{4}
\providecommand{\BIBentryALTinterwordspacing}{\spaceskip=\fontdimen2\font plus
\BIBentryALTinterwordstretchfactor\fontdimen3\font minus \fontdimen4\font\relax}
\providecommand{\BIBforeignlanguage}[2]{{%
\expandafter\ifx\csname l@#1\endcsname\relax
\typeout{** WARNING: IEEEtran.bst: No hyphenation pattern has been}%
\typeout{** loaded for the language `#1'. Using the pattern for}%
\typeout{** the default language instead.}%
\else
\language=\csname l@#1\endcsname
\fi
#2}}
\providecommand{\BIBdecl}{\relax}
\BIBdecl

\bibitem{RN1}
E.~Ie, V.~Jain, J.~Wang, S.~Narvekar, R.~Agarwal, R.~Wu, H.-T. Cheng, T.~Chandra, and C.~Boutilier, ``Slateq: a tractable decomposition for reinforcement learning with recommendation sets,'' in \emph{Proceedings of the 28th International Joint Conference on Artificial Intelligence}, ser. IJCAI'19.\hskip 1em plus 0.5em minus 0.4em\relax AAAI Press, 2019, p. 2592–2599.

\bibitem{RN16}
L.~Zou, L.~Xia, Z.~Ding, J.~Song, W.~Liu, and D.~Yin, ``Reinforcement learning to optimize long-term user engagement in recommender systems,'' in \emph{Proceedings of the 25th ACM SIGKDD International Conference on Knowledge Discovery \& Data Mining}, ser. KDD '19.\hskip 1em plus 0.5em minus 0.4em\relax New York, NY, USA: Association for Computing Machinery, 2019, p. 2810–2818.

\bibitem{RN15}
L.~Huang, M.~Fu, F.~Li, H.~Qu, Y.~Liu, and W.~Chen, ``A deep reinforcement learning based long-term recommender system,'' \emph{Knowledge-Based Systems}, vol. 213, p. 106706, 2021.

\bibitem{RN24}
X.~Zhao, C.~Gu, H.~Zhang, X.~Yang, X.~Liu, J.~Tang, and H.~Liu, ``Dear: Deep reinforcement learning for online advertising impression in recommender systems,'' in \emph{Proceedings of the AAAI conference on artificial intelligence}, ser. AAAI'21, vol.~35, no.~1, 2021, pp. 750--758.

\bibitem{RN18}
Q.~Li, Z.~Wen, Z.~Wu, S.~Hu, N.~Wang, Y.~Li, X.~Liu, and B.~He, ``A survey on federated learning systems: Vision, hype and reality for data privacy and protection,'' \emph{IEEE Trans. on Knowl. and Data Eng.}, vol.~35, no.~4, p. 3347–3366, Apr. 2023.

\bibitem{RN22}
J.~C. Duchi, M.~I. Jordan, and M.~J. Wainwright, ``Privacy aware learning,'' \emph{J. ACM}, vol.~61, no.~6, pp. 1--57, Dec. 2014.

\bibitem{RN21}
Y.~Arjevani and O.~Shamir, ``Communication complexity of distributed convex learning and optimization,'' in \emph{Proceedings of the 29th International Conference on Neural Information Processing Systems}, ser. NIPS'15.\hskip 1em plus 0.5em minus 0.4em\relax Cambridge, MA, USA: MIT Press, 2015, p. 1756–1764.

\bibitem{RN20}
T.~Kavarakuntla, L.~Han, H.~Lloyd, A.~Latham, and S.~B. Akintoye, ``Performance analysis of distributed deep learning frameworks in a multi-gpu environment,'' in \emph{The 20th International Conference on Ubiquitous Computing and Communications}, ser. IUCC'21, 2021, pp. 406--413.

\bibitem{RN19}
X.~Cao, T.~Başar, S.~Diggavi, Y.~C. Eldar, K.~B. Letaief, H.~V. Poor, and J.~Zhang, ``Communication-efficient distributed learning: An overview,'' \emph{IEEE Journal on Selected Areas in Communications}, vol.~41, no.~4, pp. 851--873, 2023.

\bibitem{DBLP:conf/aistats/McMahanMRHA17}
B.~McMahan, E.~Moore, D.~Ramage, S.~Hampson, and B.~A. y~Arcas, ``Communication-efficient learning of deep networks from decentralized data,'' in \emph{Proceedings of the 20th International Conference on Artificial Intelligence and Statistics}, ser. AISTATS'17, A.~Singh and X.~J. Zhu, Eds., vol.~54.\hskip 1em plus 0.5em minus 0.4em\relax {PMLR}, 2017, pp. 1273--1282.

\bibitem{yang2020federated}
Y.~Qiang, ``Federated recommendation systems,'' in \emph{2019 IEEE International Conference on Big Data (Big Data)}, 2019, pp. 1--1.

\bibitem{jalalirad2019simple}
A.~Jalalirad, M.~Scavuzzo, C.~Capota, and M.~Sprague, ``A simple and efficient federated recommender system,'' in \emph{Proceedings of the 6th IEEE/ACM International Conference on Big Data Computing, Applications and Technologies}, ser. BDCAT '19.\hskip 1em plus 0.5em minus 0.4em\relax New York, NY, USA: Association for Computing Machinery, 2019, p. 53–58.

\bibitem{muhammad2020fedfast}
K.~Muhammad, Q.~Wang, D.~O'Reilly-Morgan, E.~Tragos, B.~Smyth, N.~Hurley, J.~Geraci, and A.~Lawlor, ``Fedfast: Going beyond average for faster training of federated recommender systems,'' in \emph{Proceedings of the 26th ACM SIGKDD International Conference on Knowledge Discovery \& Data Mining}, ser. KDD '20.\hskip 1em plus 0.5em minus 0.4em\relax New York, NY, USA: Association for Computing Machinery, 2020, p. 1234–1242.

\bibitem{tan2020federated}
B.~Tan, B.~Liu, V.~Zheng, and Q.~Yang, ``A federated recommender system for online services,'' in \emph{Proceedings of the 14th ACM Conference on Recommender Systems}, ser. RecSys '20.\hskip 1em plus 0.5em minus 0.4em\relax New York, NY, USA: Association for Computing Machinery, 2020, p. 579–581.

\bibitem{imran2023refrs}
M.~Imran, H.~Yin, T.~Chen, Q.~V.~H. Nguyen, A.~Zhou, and K.~Zheng, ``Refrs: Resource-efficient federated recommender system for dynamic and diversified user preferences,'' \emph{ACM Trans. Inf. Syst.}, vol.~41, no.~3, pp. 1--30, Feb. 2023.

\bibitem{RN4}
Y.~Deng, M.~M. Kamani, and M.~Mahdavi, ``Adaptive personalized federated learning,'' arXiv:2003.13461 [cs.LG], 2020.

\bibitem{RN7}
H.~H. Zhuo, W.~Feng, Y.~Lin, Q.~Xu, and Q.~Yang, ``Federated deep reinforcement learning,'' arXiv:1901.08277 [cs.LG], 2020.

\bibitem{RN26}
E.~Ie, C.~wei Hsu, M.~Mladenov, V.~Jain, S.~Narvekar, J.~Wang, R.~Wu, and C.~Boutilier, ``Recsim: A configurable simulation platform for recommender systems,'' arXiv:1909.04847 [cs.LG], 2019.

\bibitem{lu2015recommender}
J.~Lu, D.~Wu, M.~Mao, W.~Wang, and G.~Zhang, ``Recommender system application developments,'' \emph{Decis. Support Syst.}, vol.~74, no.~C, p. 12–32, Jun. 2015.

\bibitem{su2009survey}
X.~Su and T.~M. Khoshgoftaar, ``A survey of collaborative filtering techniques,'' \emph{Adv. in Artif. Intell.}, vol. 2009, pp. 2--4, Jan. 2009.

\bibitem{koren2021advances}
Y.~Koren and R.~Bell, \emph{Advances in Collaborative Filtering}.\hskip 1em plus 0.5em minus 0.4em\relax Boston, MA: Springer US, 2015, pp. 77--118.

\bibitem{thorat2015survey}
P.~B. Thorat, R.~M. Goudar, and S.~Barve, ``Survey on collaborative filtering, content-based filtering and hybrid recommendation system,'' \emph{International Journal of Computer Applications}, vol. 110, no.~4, pp. 31--36, 2015.

\bibitem{wang2006unifying}
J.~Wang, A.~P. de~Vries, and M.~J.~T. Reinders, ``Unifying user-based and item-based collaborative filtering approaches by similarity fusion,'' in \emph{Proceedings of the 29th Annual International ACM SIGIR Conference on Research and Development in Information Retrieval}, ser. SIGIR '06.\hskip 1em plus 0.5em minus 0.4em\relax New York, NY, USA: Association for Computing Machinery, 2006, p. 501–508.

\bibitem{pazzani2007content}
M.~J. Pazzani and D.~Billsus, ``Content-based recommendation systems,'' \emph{The adaptive web: methods and strategies of web personalization}, pp. 325--341, 2007.

\bibitem{gershman2010decision}
A.~Gershman, A.~Meisels, K.-H. L{\"u}ke, L.~Rokach, A.~Schclar, and A.~Sturm, ``A decision tree based recommender system,'' in \emph{The 10th International Conferenceon Innovative Internet Community Systems}, ser. I2CS.\hskip 1em plus 0.5em minus 0.4em\relax Gesellschaft f{\"u}r Informatik eV, 2010, pp. 170--179.

\bibitem{oku2006context}
K.~Oku, S.~Nakajima, J.~Miyazaki, and S.~Uemura, ``Context-aware svm for context-dependent information recommendation,'' in \emph{Proceedings of the 7th International Conference on Mobile Data Management}, ser. MDM '06.\hskip 1em plus 0.5em minus 0.4em\relax USA: IEEE Computer Society, 2006, p. 109.

\bibitem{gupta2020architectural}
U.~Gupta, C.-J. Wu, X.~Wang, M.~Naumov, B.~Reagen, D.~Brooks, B.~Cottel, K.~Hazelwood, M.~Hempstead, B.~Jia, H.-H.~S. Lee, A.~Malevich, D.~Mudigere, M.~Smelyanskiy, L.~Xiong, and X.~Zhang, ``The architectural implications of facebook's dnn-based personalized recommendation,'' in \emph{IEEE International Symposium on High Performance Computer Architecture}, ser. HPCA, 2020, pp. 488--501.

\bibitem{RN27}
X.~Chen, L.~Yao, J.~McAuley, G.~Zhou, and X.~Wang, ``A survey of deep reinforcement learning in recommender systems: A systematic review and future directions,'' arXiv:2109.03540 [cs.IR], 2021.

\bibitem{RN28}
M.~M. Afsar, T.~Crump, and B.~Far, ``Reinforcement learning based recommender systems: A survey,'' \emph{ACM Comput. Surv.}, vol.~55, no.~7, pp. 1--38, Dec. 2022.

\bibitem{tan2023adaptive}
X.~Tan, Y.~Deng, C.~Qu, S.~Xue, X.~Shi, J.~Zhang, and X.~Qiu, ``Adaptive learning on user segmentation: Universal to specific representation via bipartite neural interaction,'' in \emph{Proceedings of the Annual International ACM SIGIR Conference on Research and Development in Information Retrieval in the Asia Pacific Region}, ser. SIGIR-AP '23.\hskip 1em plus 0.5em minus 0.4em\relax New York, NY, USA: Association for Computing Machinery, 2023, p. 205–211.

\bibitem{9904958}
X.~Wang, S.~Wang, X.~Liang, D.~Zhao, J.~Huang, X.~Xu, B.~Dai, and Q.~Miao, ``Deep reinforcement learning: A survey,'' \emph{IEEE Transactions on Neural Networks and Learning Systems}, pp. 1--15, 2022.

\bibitem{kaloev2021experiments}
M.~Kaloev and G.~Krastev, ``Experiments focused on exploration in deep reinforcement learning,'' in \emph{2021 5th International Symposium on Multidisciplinary Studies and Innovative Technologies (ISMSIT)}.\hskip 1em plus 0.5em minus 0.4em\relax IEEE, 2021, pp. 351--355.

\bibitem{deng2016deep}
Y.~Deng, F.~Bao, Y.~Kong, Z.~Ren, and Q.~Dai, ``Deep direct reinforcement learning for financial signal representation and trading,'' \emph{IEEE transactions on neural networks and learning systems}, vol.~28, no.~3, pp. 653--664, 2016.

\bibitem{QIU2022107689}
X.~Qiu, X.~Tan, Q.~Li, S.~Chen, Y.~Ru, and Y.~Jin, ``A latent batch-constrained deep reinforcement learning approach for precision dosing clinical decision support,'' \emph{Knowledge-based systems}, vol. 237, p. 107689, 2022.

\bibitem{CHEN202247}
S.~Chen, X.~Qiu, X.~Tan, Z.~Fang, and Y.~Jin, ``A model-based hybrid soft actor-critic deep reinforcement learning algorithm for optimal ventilator settings,'' \emph{Inf. Sci.}, vol. 611, no.~C, p. 47–64, Sep. 2022.

\bibitem{RN29}
G.~Zheng, F.~Zhang, Z.~Zheng, Y.~Xiang, N.~J. Yuan, X.~Xie, and Z.~Li, ``Drn: A deep reinforcement learning framework for news recommendation,'' in \emph{Proceedings of the 2018 World Wide Web Conference}, ser. WWW '18.\hskip 1em plus 0.5em minus 0.4em\relax Republic and Canton of Geneva, CHE: International World Wide Web Conferences Steering Committee, 2018, p. 167–176.

\bibitem{RN30}
Y.~Lei, Z.~Wang, W.~Li, and H.~Pei, ``Social attentive deep q-network for recommendation,'' in \emph{Proceedings of the 42nd International ACM SIGIR Conference on Research and Development in Information Retrieval}, ser. SIGIR'19.\hskip 1em plus 0.5em minus 0.4em\relax New York, NY, USA: Association for Computing Machinery, 2019, p. 1189–1192.

\bibitem{RN32}
M.~Chen, A.~Beutel, P.~Covington, S.~Jain, F.~Belletti, and E.~H. Chi, ``Top-k off-policy correction for a reinforce recommender system,'' in \emph{Proceedings of the Twelfth ACM International Conference on Web Search and Data Mining}, ser. WSDM '19.\hskip 1em plus 0.5em minus 0.4em\relax New York, NY, USA: Association for Computing Machinery, 2019, p. 456–464.

\bibitem{RN8}
H.~Zhu, J.~Xu, S.~Liu, and Y.~Jin, ``Federated learning on non-iid data: A survey,'' \emph{Neurocomput.}, vol. 465, no.~C, p. 371–390, Nov. 2021.

\bibitem{sattler2019robust}
F.~Sattler, S.~Wiedemann, K.-R. Müller, and W.~Samek, ``Robust and communication-efficient federated learning from non-i.i.d. data,'' \emph{IEEE Transactions on Neural Networks and Learning Systems}, vol.~31, no.~9, pp. 3400--3413, 2020.

\bibitem{RN35}
H.~Wang, Z.~Kaplan, D.~Niu, and B.~Li, ``Optimizing federated learning on non-iid data with reinforcement learning,'' in \emph{IEEE INFOCOM 2020 - IEEE Conference on Computer Communications}, 2020, pp. 1698--1707.

\bibitem{RN36}
P.~Zhang, C.~Wang, C.~Jiang, and Z.~Han, ``Deep reinforcement learning assisted federated learning algorithm for data management of iiot,'' \emph{IEEE Transactions on Industrial Informatics}, vol.~17, no.~12, pp. 8475--8484, 2021.

\bibitem{RN38}
M.~Ahmadi, A.~Taghavirashidizadeh, D.~Javaheri, A.~Masoumian, S.~{Jafarzadeh Ghoushchi}, and Y.~Pourasad, ``Dqre-scnet: A novel hybrid approach for selecting users in federated learning with deep-q-reinforcement learning based on spectral clustering,'' \emph{Journal of King Saud University - Computer and Information Sciences}, vol.~34, no.~9, pp. 7445--7458, 2022.

\bibitem{RN37}
Y.~Zhan, P.~Li, and S.~Guo, ``Experience-driven computational resource allocation of federated learning by deep reinforcement learning,'' in \emph{2020 IEEE International Parallel and Distributed Processing Symposium (IPDPS)}, 2020, pp. 234--243.

\bibitem{louviere2000stated}
J.~J. Louviere, D.~A. Hensher, and J.~D. Swait, \emph{Stated choice methods: analysis and application}.\hskip 1em plus 0.5em minus 0.4em\relax USA: Cambridge University Press, 2001.

\bibitem{DBLP:conf/kdd/Joachims02}
T.~Joachims, ``Optimizing search engines using clickthrough data,'' in \emph{Proceedings of the Eighth ACM SIGKDD International Conference on Knowledge Discovery and Data Mining}, ser. KDD '02.\hskip 1em plus 0.5em minus 0.4em\relax New York, NY, USA: Association for Computing Machinery, 2002, p. 133–142.

\bibitem{DBLP:conf/wsdm/CraswellZTR08}
N.~Craswell, O.~Zoeter, M.~Taylor, and B.~Ramsey, ``An experimental comparison of click position-bias models,'' in \emph{Proceedings of the 2008 International Conference on Web Search and Data Mining}, ser. WSDM '08.\hskip 1em plus 0.5em minus 0.4em\relax New York, NY, USA: Association for Computing Machinery, 2008, p. 87–94.

\bibitem{wang2022individual}
L.~Wang, Y.~Zhang, Y.~Hu, W.~Wang, C.~Zhang, Y.~Gao, J.~Hao, T.~Lv, and C.~Fan, ``Individual reward assisted multi-agent reinforcement learning,'' in \emph{Proceedings of the 39th International Conference on Machine Learning}, ser. PMLR, K.~Chaudhuri, S.~Jegelka, L.~Song, C.~Szepesvari, G.~Niu, and S.~Sabato, Eds., vol. 162, 17--23 Jul 2022, pp. 23\,417--23\,432.

\bibitem{DBLP:journals/corr/abs-2303-04689}
D.~Neumann, A.~Lutz, K.~M\"{u}ller, and W.~Samek, ``A privacy preserving system for movie recommendations using federated learning,'' \emph{ACM Trans. Recomm. Syst.}, vol.~3, no.~2, pp. 1--51, Nov. 2024.

\bibitem{misra2019mish}
D.~Misra, ``Mish: A self regularized non-monotonic activation function,'' in \emph{British Machine Vision Conference}, ser. BMVC'20, 2020.

\bibitem{DBLP:journals/tce/JaveedSKJII24}
D.~Javeed, M.~S. Saeed, P.~Kumar, A.~Jolfaei, S.~Islam, and A.~K. M.~N. Islam, ``Federated learning-based personalized recommendation systems: An overview on security and privacy challenges,'' \emph{IEEE Transactions on Consumer Electronics}, vol.~70, no.~1, pp. 2618--2627, 2024.

\bibitem{DBLP:journals/ojcomps/ChronisVHSANBD24}
C.~Chronis, I.~Varlamis, Y.~Himeur, A.~N. Sayed, T.~M. AL-Hasan, A.~Nhlabatsi, F.~Bensaali, and G.~Dimitrakopoulos, ``A survey on the use of federated learning in privacy-preserving recommender systems,'' \emph{IEEE Open Journal of the Computer Society}, vol.~5, pp. 227--247, 2024.

\end{thebibliography}

\section*{Biography Section}

 





\begin{IEEEbiography}[{\includegraphics[width=1in,height=1.25in,clip,keepaspectratio]{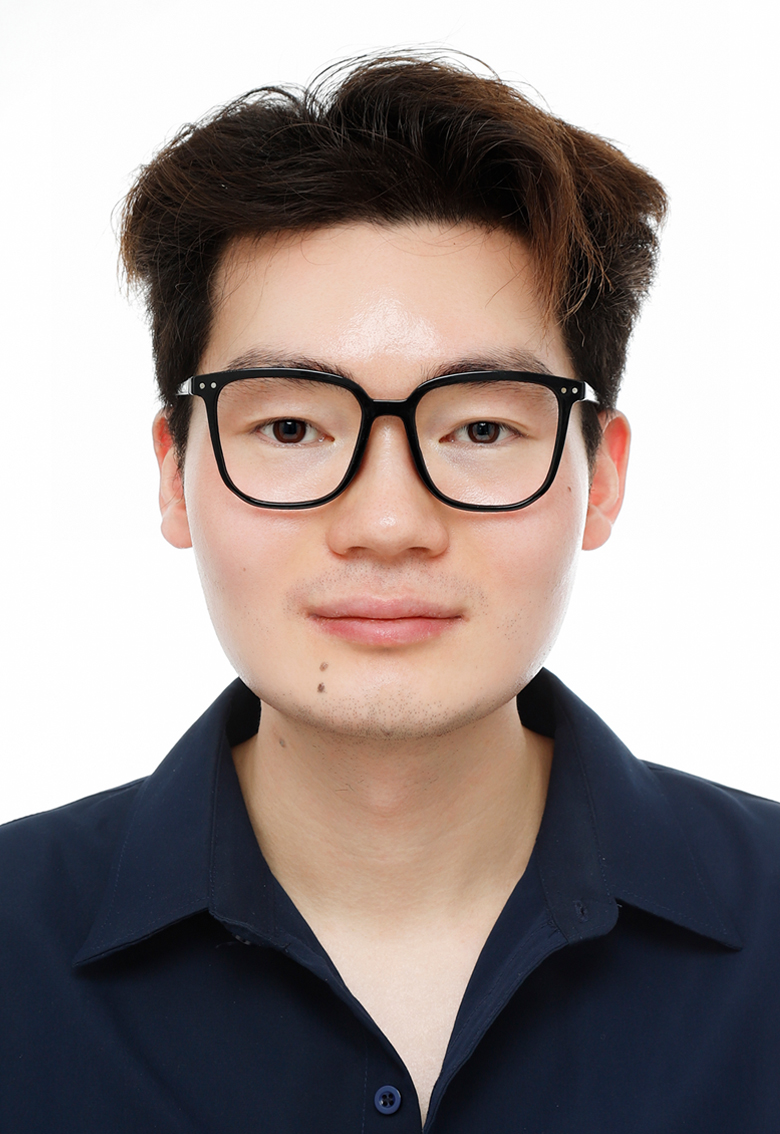}}]{Yongxin Deng}
is a graduate student at the School of Electronic and Electrical Engineering, Shanghai University of Engineering Science, Shanghai, China. His research interests include artificial intelligence and healthcare informatics.
\end{IEEEbiography}

\begin{IEEEbiography}[{\includegraphics[width=1in,height=1.25in,clip,keepaspectratio]{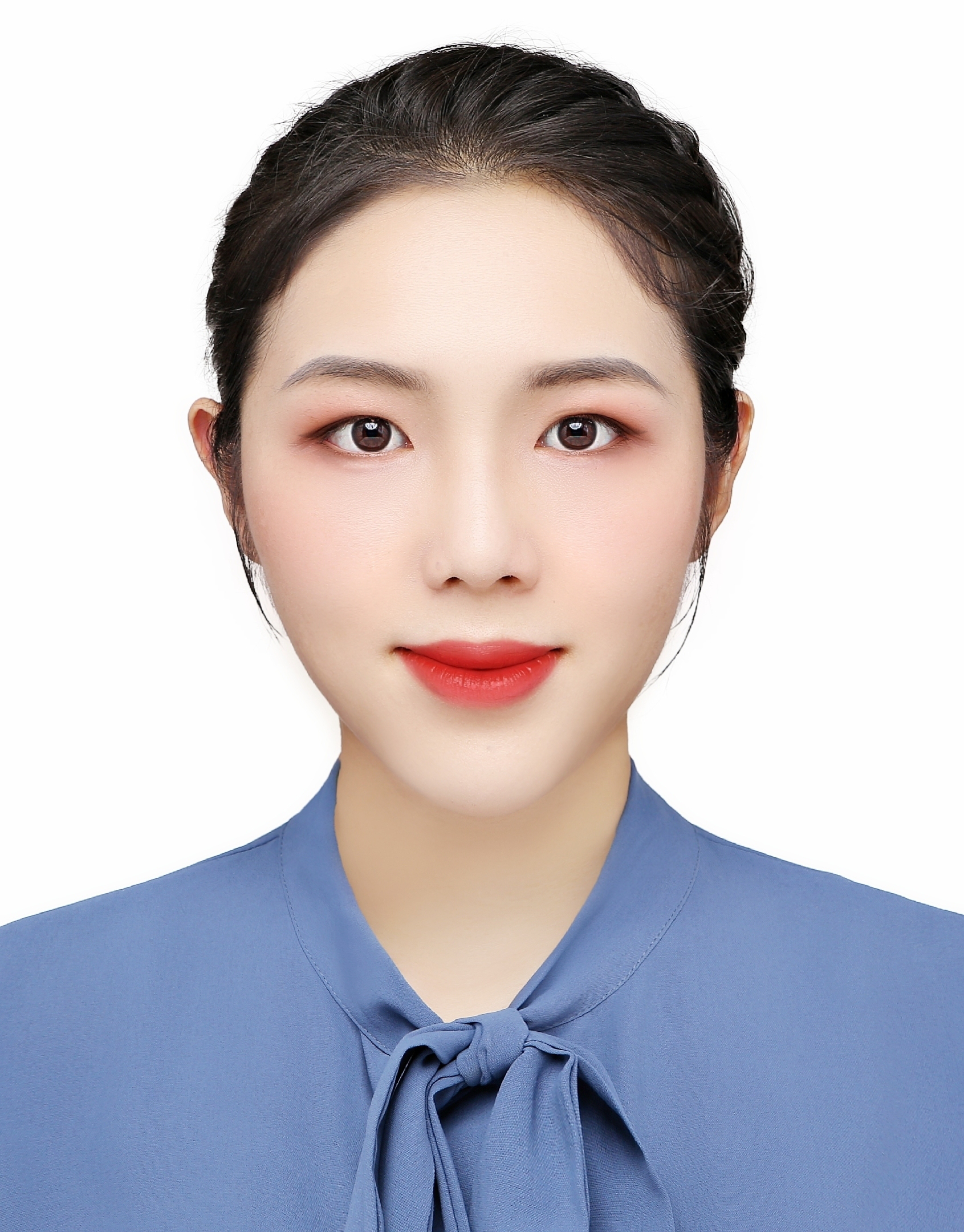}}]{Xihe Qiu}
is an Associate Professor at the School of Electronic and Electrical Engineering, Shanghai University of Engineering Science, Shanghai, China. His research focuses on machine learning and clinical decision support systems. Email: qiuxihe@sues.edu.cn
\end{IEEEbiography}

\begin{IEEEbiography}[{\includegraphics[width=1in,height=1.25in,clip,keepaspectratio]{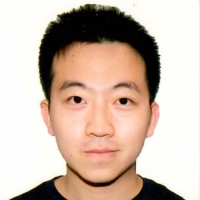}}]{Xiaoyu Tan}
received the B.E. degree in Mechanical Engineering from Northeastern University, China, in 2015, and the Ph.D. degree from the National University of Singapore, Singapore, in 2019. He worked as an Algorithm Engineer at Ant Group from 2019 to 2022. He is currently an AI Algorithm Engineer with INF Tech, Shanghai, China. His research interests include artificial intelligence and machine learning algorithms.
\end{IEEEbiography}

\begin{IEEEbiography}[{\includegraphics[width=1in,height=1.25in,clip,keepaspectratio]{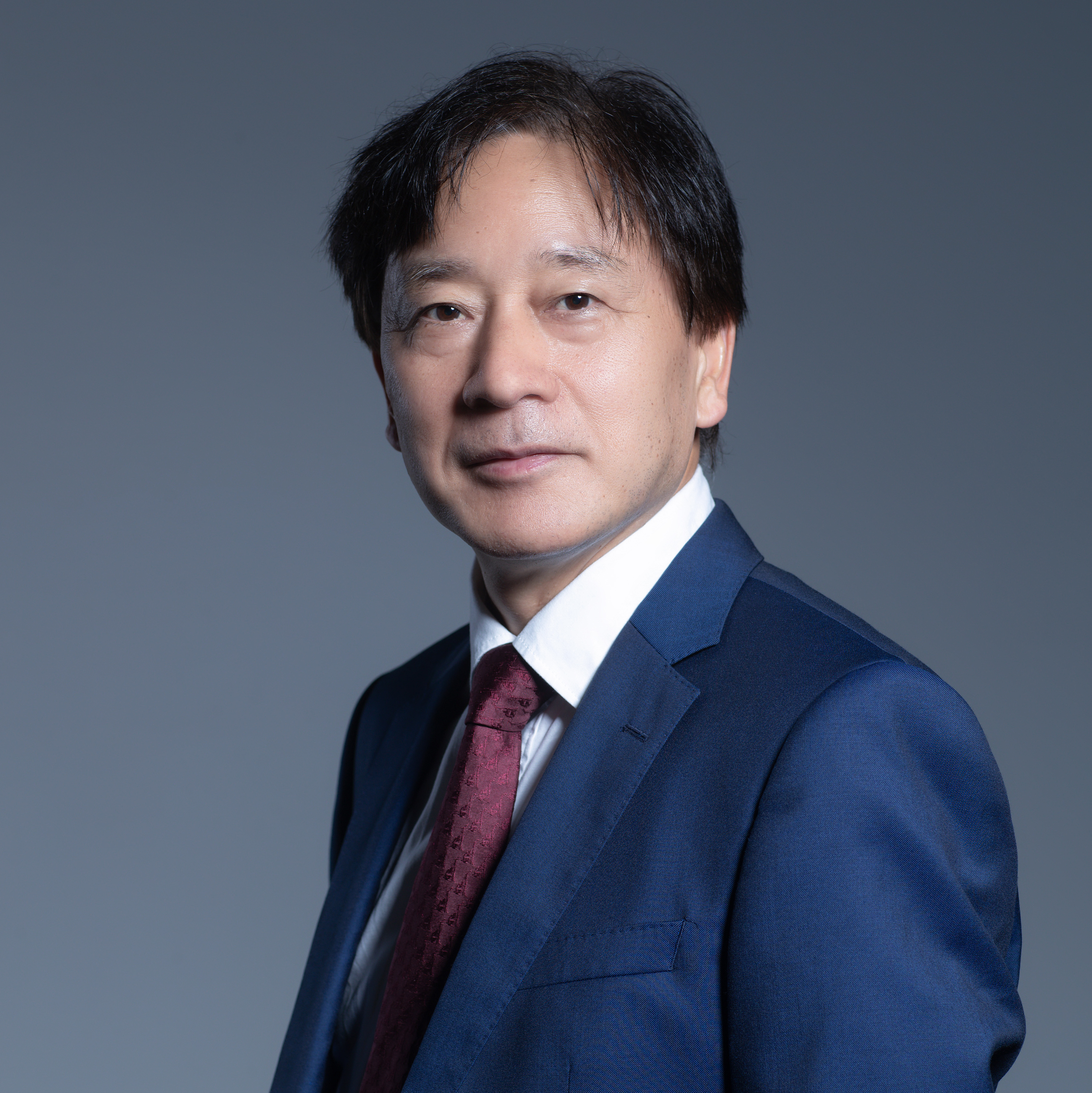}}]{Yaochu Jin (Fellow, IEEE)} 
received the BSc, MSc, and PhD degrees in automatic control from Zhejiang University, Hangzhou, China, in 1988, 1991, and 1996, respectively, and the Dr.-Ing. degree from Ruhr-University Bochum, Bochum, Germany, in 2001. He is presently Chair Professor of AI with the School of Engineering, Westlake University, Hangzhou, China. He is the Head of the Department of Artificial Engineering and leads the Trustworthy and General AI Laboratory. He was Alexander von Humboldt Professor for AI with the Faculty of Technology, Bielefeld University, Germany, and Surrey Distinguished Chair of Computational Intelligence, University of Surrey, U.K. He was a Finland Distinguished Professor and a Changjiang Distinguished Visiting Professor in China. His main research interests include data-driven evolutionary optimization, trustworthy machine learning, multiobjective evolutionary learning, and evolutionary developmental systems. Dr Jin is the recipient of the 2025 IEEE Frank Rosenblatt Award, the 2018, 2021 and 2024 IEEE Transactions on Evolutionary Computation Outstanding Paper Award, and the 2015, 2017, and 2020 IEEE Computational Intelligence Magazine Outstanding Paper Award. He is currently the President of the IEEE Computational Intelligence Society and Editor-in-Chief of Complex \& Intelligent Systems. He is a member of Academia Europaea.

\end{IEEEbiography}

\vfill

\end{document}